\documentclass[prb,twocolumn]{revtex4-1}
\usepackage{times}
\usepackage{bbm}
\usepackage[dvips]{graphicx}
\usepackage{latexsym,amsmath,amssymb,bm,euscript,eufrak}
\usepackage[dvips]{color}
\usepackage{multirow}
\usepackage{calligra}
\usepackage{color}
\usepackage[colorlinks=true,linkcolor=blue,citecolor=blue]{hyperref}
\usepackage{hyperref}

\newcommand{\scripty}[1]{w}

\def\ra{\rangle} 
\def\la{\langle} 

\def\ztwo{\mathbb{Z}_2}
\def\ztwot{\mathbb{Z}_2^T}
\newcommand{\cp}{$CP^1$ }
\newcommand{\nccp}{$NCCP^1$ }

\def\up{\uparrow}
\def\dn{\downarrow}

\def\cJ{{J}}
\def\cQ{{Q}}
\def\cP{{P}}

\def\cG{{G}}
\def\Aext{A^{\rm ext}}

\def\hext{h^{\rm ext}}
\def\uu{B}
\def\MM{N}
\def\mm{n}

\begin{document}

\title{Model Realization and Numerical Studies of a Three-Dimensional Bosonic Topological Insulator and Symmetry-Enriched Topological Phases}
\date{\today}

\author{Scott D. Geraedts}
\author{Olexei I. Motrunich}
\affiliation{Department of Physics, California Institute of Technology, Pasadena, California 91125, USA}

\begin{abstract}
We study a topological phase of interacting bosons in (3+1) dimensions which is protected by charge conservation and time-reversal symmetry. We present an explicit lattice model which realizes this phase and which can be studied in sign-free Monte Carlo simulations. The idea behind our model is to bind bosons to topological defects called hedgehogs. We determine the phase diagram of the model and identify a phase where such bound states are proliferated.  In this phase we observe a Witten effect in the bulk whereby an external monopole binds half of the elementary boson charge, which confirms that it is a bosonic topological insulator. We also study the boundary between the topological insulator and a trivial insulator. We find a surface phase diagram which includes exotic superfluids, a topologically ordered phase, and a phase with a Hall effect quantized to one-half of the value possible in a purely two-dimensional system. We also present models that realize symmetry-enriched topologically-ordered phases by binding multiple hedgehogs to each boson; these phases show charge fractionalization and intrinsic topological order as well as a fractional Witten effect.
\end{abstract}
\maketitle

\section{Introduction}
The study of topological phases of matter has been a major component of condensed matter research in recent decades. Among the many phases studied, the topological insulator (TI) is one of the most prominent.\cite{KaneHasanRMP,QiZhangRMP} The TI is a three-dimensional phase of free fermions. Though it is insulating in the bulk, its topological behavior can be deduced from its unusual surface properties, in particular the odd number of Dirac cones it has on its surface. The topological insulator is an example of a symmetry-protected topological phase (SPT).  Like all SPT's, it has short-ranged entanglement, which implies that it has only conventional excitations in the bulk and a unique ground state on any closed manifold. This is in contrast to intrinsically topologically ordered states like the fractional quantum Hall states. The relevant symmetries for the topological insulator are charge conservation and time-reversal, and if either of these symmetries is violated, the phase loses its topological properties.

One obvious extension of research into topological insulators is to consider the effects of interactions on their properties. This is however a difficult task. Many of the methods used to study TI's involve the properties of their band structure, and these methods obviously do not apply if interactions are strong. As an introduction to this difficult problem, one can try to study an analog of the topological insulator, constructed of interacting {\em bosons} instead of fermions. In bosonic systems we know that the non-interacting case would be a condensate, so we can be sure that the topological behavior is due to the interactions. In addition, certain theoretical techniques, like the Monte Carlo studies employed in this paper, work mainly for bosonic systems.

The study of topological phases of interacting bosons is relatively recent, but much progress has been made.\cite{WenScience,*WenPRB,LuVishwanath,SenthilLevin,SenthilVishwanath,FQHE,TurnerVishwanath,SenthilReview,BiRasmussenXu, WangSenthil,Kapustin2014,Wen2014} 
Chen, Liu, Gu, and Wen\cite{WenScience,*WenPRB} have used group cohomology theory to determine which symmetries and dimensions can lead to non-trivial topological phases. However, this approach tells us little about the properties of these phases, which must be determined through other methods. One well-studied case is that of SPT phases with $U(1)$ symmetry in two dimensions. Lu and Vishwanath\cite{LuVishwanath} proposed a phenomenological Chern-Simons field theory to describe both the bulk and edges of these states and showed that such states have a Hall effect quantized to an even integer (in units of $e^2/h$) and possess gapless counter-propagating edge modes. Therefore such states are called ``bosonic integer quantum Hall phases''. Senthil and Levin\cite{SenthilLevin} proposed how to realize such phases in a quantum Hall-like setting by starting with two species of bosons in a magnetic field and using mutual flux attachment.  One drawback of the flux attachment technique is that it is difficult to relate it precisely to a microscopic model.  To address this, in our work\cite{FQHE} we provided an alternative exactly solvable model that replaces flux attachment by a (precisely formulated) dynamical binding of bosons to vortices.  We studied this model in Monte Carlo and confirmed that it realizes the integer Quantum Hall phase of bosons, including observation of the gapless edge modes.  We also showed that by binding multiple vortices to bosons we can realize Symmetry-Enriched Topological phases (SET)\cite{EssinHermele,MesarosRan} with long-ranged entanglement.

In this paper we focus on understanding interacting bosonic topological phases which are analogs of the electronic topological insulator in three dimensions. Motivated by the formal cohomology results,\cite{WenScience,*WenPRB} Vishwanath and Senthil in an inspirational paper\cite{SenthilVishwanath} found effective field theories which can describe both the bulk and the surface of a three-dimensional ``bosonic topological insulator'' with charge conservation and time-reversal symmetry. They found exotic behavior on the surface which can be used to assert the topological behavior in the bulk.

In particular, Vishwanath and Senthil found three kinds of exotic phases on the surface of the bosonic TI. These surface phases cannot exist in a purely two-dimensional system, and their existence on the surface of a three-dimensional system shows that the system is topological. The first kind of phase is a superfluid, which spontaneously breaks charge conservation symmetry. There are actually several different types of superfluids which can exist on the surface. The gapped vortex excitations in these superfluids have properties which cannot exist in a purely two-dimensional system.  Phase transitions between the different superfluids are predicted to be deconfined critical points. Another kind of surface phase appears when we break time-reversal symmetry on the surface. This phase has a Hall conductivity quantized to one-half of the elementary value possible in a purely two-dimensional bosonic system. Since the Hall conductivity in the bosonic integer quantum Hall effect is quantized to even integers, this surface phase is expected to have an odd integer Hall conductivity. Finally, it is possible to have a surface phase which breaks no symmetries but has a symmetry-enriched intrinsic topological order of a kind impossible in a purely two-dimensional system with these symmetries.

In both the two- and three-dimensional cases, the topological behavior can be thought of as coming from the binding of bosons to point topological defects, and the condensation of such objects. In our construction in two dimensions the topological defects were vortices, and the binding is essentially an exact dynamical realization of the flux attachment. In the three-dimensional systems that we will study here, the topological defects are hedgehogs which we introduce by adding an additional $SO(3)$ ``spin'' symmetry, and we construct a bosonic TI by binding hedgehogs of this symmetry to the bosons.\cite{SenthilVishwanath}
In an independent work, Metlitski and Fisher also produced a construction which explicitly binds hedgehog-like ``monopole'' objects to bosons without enlarging the continuous symmetry and showed that it gives the three-dimensional bosonic TI,\cite{Max} while the present setting is a bit simpler to analyze and is amenable to Monte Carlo studies.

In this work, we construct explicit models which realize the interacting bosonic analog of the topological insulator. The models have both charge conservation $U(1)$ symmetry and a time reversal $\ztwot$ symmetry to be discussed in the main text. We present two different models which realize this physics. In the first model, described in Section \ref{section::Heisenberg}, the spins are represented by $SO(3)$ degrees of freedom with Heisenberg interactions. We introduce a term in our action which energetically binds hedgehogs to bosons, and we show that this term can lead to a phase (which we call the ``binding phase'') where these bound states are proliferated.

In order to show that the binding phase is indeed the bosonic topological insulator, we will attempt to find the phases predicted by Vishwanath and Senthil on its surface. Initially we find only superfluids on the surface. In our Monte Carlo simulations, we do not have simple access to properties of the gapped vortex excitations in these surface phases. Therefore we cannot determine whether the superfluids we observe are the exotic superfluids predicted by Vishwanath and Senthil, or more conventional superfluids. We find that the surface superfluids in our model are connected by a direct transition. If this transition were second-order, then it could be the predicted deconfined critical point. However, we cannot access large enough system sizes to determine the order of the observed phase transition.

We can try to find other exotic surface phases by explicitly breaking the $\ztwot$ symmetry of our model on the surface. If the bulk of the system is in a topological phase, this will lead to a Hall conductivity quantized to an odd integer. In order to measure this Hall conductivity in our system, we need to introduce external gauge fields. In the model in Section \ref{section::Heisenberg}, this is complicated by our definition of the hedgehog number, which is a discontinuous function of the Heisenberg spins. In this case, we do not know how to properly couple to the external gauge fields, which prevents us from measuring the surface Hall conductivity.

To remedy this, in Section~\ref{section::CP1} we introduce a second model which binds hedgehogs to bosons. In this model, we represent the spins with an easy-plane \cp model. We can identify hedgehogs with the monopoles of the internal compact gauge field of this \cp model. This formulation will allow us to make several measurements which demonstrate that our phase is a bosonic topological insulator.

One measurement that we can make is called the Witten effect.\cite{MaxWitten, Max, YeWen2014, YeWang2014} This is the binding of one-half of a boson charge to external monopoles in the spin sector. We introduce such external monopoles into the bulk of our system and find that each binds precisely half of a boson charge.

We then determine the surface phases of our model. We again find superfluids connected by a direct transition.  We can also break the $\ztwot$ symmetry on the surface. In the \cp model, we can measure the surface Hall conductivity and we find it to be quantized to odd integers as predicted. We reiterate that this Hall conductivity cannot be observed in a purely two-dimensional model, and so we must be measuring the Hall conductivity on the surface of a topological phase. Finally, we also find a surface phase which breaks none of the symmetries of the model. We suspect that this phase has symmetry-enriched intrinsic topological order of the kind predicted by Vishwanath and Senthil,\cite{SenthilVishwanath} but we do not know how to test this using Monte Carlo.

In our studies of the (2+1)-dimensional ``bosonic quantum Hall'' phases,\cite{FQHE} we found that we can obtain symmetry-enriched phases with intrinsic topological order by binding multiple topological defects (vortices) to bosons. In Section \ref{section::multiple} we consider such a binding in (3+1) dimensions. We find that binding of multiple hedgehogs to bosons leads to a bulk phase with intrinsic topological order, thus opening the study of bosonic SET phases in (3+1) dimensions. To help understand the properties of this SET phase, in the Appendix we consider a more abstract model where monopoles of a compact quantum electrodynamics (CQED) are bound to bosons; we analyze the resulting SPT-like and SET-like phases in such a CQED$\times$boson theory, and also discuss how things change upon including matter fields coupled to the CQED, which is the situation in the \cp representation of the spins.

\section{Realizing the topological insulator by binding bosons to hedgehogs of $SO(3)$ spins}
\label{section::Heisenberg}

We first study the binding between bosons and topological defects by using hedgehogs of a Heisenberg model as our topological defects. We demonstrate the existence of a phase which can be loosely viewed as a condensate of bound states of bosons and hedgehogs, and explore its properties. This model provides an intuitive introduction to the physics of such binding. Additional properties of the resulting SPT phase will be considered in Section~\ref{section::CP1}.

\subsection{Model and its Bulk Phase Diagram}
\label{subsec::bulkheis}
We study the following action, in (3+1)D Euclidean space-time:
\begin{equation}
S=S_{\rm spin}+\frac{\lambda}{2}\sum_{r,\mu} [ J_\mu(r)- Q_\mu(r)]^2.
\label{action}
\end{equation}
$S_{\rm spin}$ is an action which controls fluctuations in the $SO(3)$ spins. The second term provides the binding interaction between bosons and hedgehogs. The bosons are represented by integer-valued conserved currents, $J_\mu(r)$, defined on the links of a four-dimensional cubic lattice, where $r$ is a site label on the lattice and $\mu\in (x,y,z,\tau)$ is a direction. These currents represent the world-lines of the bosons in the (3+1)-dimensional space-time. The $Q_\mu(r)$ variables represent the hedgehogs, which are also integer-valued conserved currents and will be defined shortly.  When the real number $\lambda$ is large, this term will bind bosons and hedgehogs together. We work with periodic boundary conditions and require no net boson charge and no net boson spatial currents, so that the $J_\mu$ space-time currents have zero total winding number; such conditions are automatically satisfied for the hedgehog currents $Q_\mu$ defined below.  Imposing such conditions on the boson currents is just a convenient choice, which does not affect the bulk physics, but allows a precise change of variables involving both $J$ and $Q$ currents [see Eq.~(\ref{shift}) below].

In this section we represent the spin degrees of freedom by $SO(3)$ unit vectors. For $S_{\rm spin}$, we take the following action, which describes a Heisenberg model:
\begin{equation}
S_{\rm spin}=-\beta\sum_{R,\mu} \vec{n}(R)\cdot \vec{n}(R+\hat{\mu}).
\label{SHeis}
\end{equation}
Here $\vec{n}(R)=(n_a,n_b,n_c)(R)$ are three-component unit vectors which represent the spins. They reside on a different lattice from the $J_\mu(r)$ variables above. This lattice has its sites labelled by $R$ and located at the centers of the (hyper)cubes of the $r$-lattice in Eq.~(\ref{action}), i.e., the $R$ and $r$ lattices are dual to each other.

From these $\vec{n}(R)$, we can define the hedgehog currents $Q_\mu(r)$ using the prescription in the literature\cite{KamalMurthy,LauDasgupta,SachdevHedgehogs,LesikAshvin} generalized to four dimensions. We summarize this prescription here. We first define variables $\alpha_\mu(R)$, which reside on links connecting the spins $\vec{n}_i\equiv\vec n(R)$ and $\vec{n}_j\equiv\vec n(R+\hat\mu)$:
\begin{equation}
e^{i\alpha_{\mu}(R)}=\frac{1+\vec{n}_i\cdot\vec{n}_j+\vec{n}_i\cdot\vec{N}_0+\vec{n}_j\cdot\vec{N}_0+i\vec{N}_0 \cdot(\vec{n}_i\times\vec{n}_j)}{\sqrt{2(1+\vec{n}_i\cdot\vec{n}_j)(1+\vec{n}_i\cdot\vec{N}_0)(1+\vec{n}_j\cdot\vec{N}_0})}.
\label{alpha}
\end{equation}
Here $\vec{N}_0$ is a reference vector which we can choose arbitrarily. 
We then define placket ``fluxes'' $\omega_{\mu\nu}\in (-\pi,\pi]$ as follows:
\begin{equation}
e^{i \omega_{\mu\nu}(R)} = 
e^{i [\nabla_\mu\alpha_\nu(R)-\nabla_\nu\alpha_\mu(R)]}.
\label{omega}
\end{equation}
Here $\nabla_\mu \alpha_\nu(R)\equiv \alpha_\nu(R+\hat{\mu})-\alpha_\nu(R)$. One can show that changing the reference vector $\vec{N}_0$ corresponds to a gauge transformation of the $\alpha_\mu(R)$ variables so that $\omega_{\mu\nu}(R)$ are independent of the reference vector. Finally, we can define the hedgehog current:
\begin{equation}
Q_\mu(r)=\frac{1}{4\pi}\epsilon_{\mu\nu\rho\sigma}\nabla_{\nu} \omega_{\rho\sigma},
\label{monopoledef}
\end{equation}
with implied summation over repeated indices.
Consider for example 
\begin{eqnarray}
Q_\tau[r=R+(\frac{1}{2},\frac{1}{2},\frac{1}{2},-\frac{1}{2})]=\frac{\nabla_x\omega_{yz}+\nabla_y\omega_{zx}+\nabla_z\omega_{xy}}{2\pi}. \nonumber
\end{eqnarray}
The right-hand-side is defined on a cube $[R, R+\hat{x}, R+\hat{y}, R+\hat{z}]$, and can also be associated with a point $R + (\hat{x} + \hat{y} + \hat{z})/2$ in the center of this cube.
The $\omega$ variables are fluxes passing through the plaquettes of this cube, and the net flux out of the cube is guaranteed to be integer multiple of $2\pi$.  We then define hedgehog number to be the net outgoing flux divided by $2\pi$.
When all four dimensions are considered, the center of this cube can be equivalently associated with a link in the $\tau$ direction from a dual lattice site $r = R + (\hat{x} + \hat{y} + \hat{z} - \hat{\tau})/2$, and the hedgehog number becomes the $\tau$-component, $Q_\tau(r)$, of the hedgehog four-current.

We have studied the above action in Monte Carlo simulations. In addition to simple independent updates of $\vec{n}$ and $J_\mu$, when $\lambda$ is large one needs to try updates which change both $J_\mu$ and $Q_\mu$ in such a way that the second term in Eq.~(\ref{action}) is unchanged. To do this we choose a spin and an amount to update it, check to see if any $Q_\mu$ variables will change, and include matching changes in the $J_\mu$ variables as part of the attempted Monte Carlo move.

In our simulations, we monitor the ``internal energy per site,'' $\epsilon= S /{\rm Vol}$, where ${\rm Vol}\equiv L^4$ is the volume of the system, which we take to have linear size $L$ in all four directions. From this, we can determine the specific heat per site:
\begin{equation}
C=(\la \epsilon^2\ra-\la\epsilon\ra^2)\times{\rm Vol}.
\end{equation}
We can locate phase transitions in our model by looking for singularities in the specific heat. We also monitor the magnetization per spin:
\begin{equation}
m = \frac{\left\la \left|\sum_R \vec{n}(R) \right|\right\ra}{\rm Vol}.
\end{equation}
When the spins are disordered the magnetization is proportional to $1/\sqrt{\rm Vol}$, while in the ordered phase the magnetization remains non-zero in the thermodynamic limit. Therefore we can use measurements of the magnetization at different sizes to determine if the spins are ordered.

To study the behavior of the boson currents, we monitor current-current correlators, defined as:
\begin{equation}
\rho_{J}({k})=\la J_\mu({k})J_\mu(-{k})\ra ~,
\label{rho}
\end{equation}
where $k$ is a wave vector, $\mu$ is a fixed direction, and 
\begin{equation}
J_\mu({k})\equiv\frac{1}{\rm \sqrt{Vol}}\sum_r J_\mu(r)e^{-i{k}\cdot r}.
\end{equation}
In the space-time isotropic system, $\rho_J({k})$ is independent of the direction $\mu$, and when we show numerical data we average over all directions to improve statistics. In an ensemble which would allow non-zero total winding number, $\rho_J(0)$ would be the familiar superfluid stiffness. In our model $J(k=0)=0$, so this measurment is not available in our simulations. Instead, we evaluate the correlators at the smallest non-zero ${k}$.  For example, if $\mu$ is in the $x$ direction, we can take ${k}_{\rm min}=(0,\frac{2\pi}{L},0,0)$, $(0,0,\frac{2\pi}{L},0)$, and $(0,0,0,\frac{2\pi}{L})$ and average over these; we exclude $(\frac{2\pi}{L},0,0,0)$ because in our ensemble the net winding of the $J_x$ current is zero, so the $J_x(k)$ evaluated at this wavevector is also zero. We also monitor current-current correlators of the hedgehog currents, $\rho_Q(k)$, which are defined in the same way as for the boson currents.

In the phase where the $J_\mu$ are gapped, only small loops contribute to the current-current correlators and $\rho_J(k_{\rm min})\sim {k}_{\rm min}^2 \sim 1/L^2$, while when the $J_\mu$ proliferate $\rho_J$ is independent of the system size. Therefore we can use finite-size scaling of this quantity to determine the locations of phase transitions. For the hedgehog currents, $\rho_Q(k_{\rm min})\sim {k}_{\rm min}^2$ in all phases, so we cannot use finite-size scaling of this quantity to find phase transitions; this is similar to the properties of vortex currents in a (2+1)-dimensional XY model and originates from effective long-range interactions of these topological defects.

\begin{figure}
\includegraphics[angle=-90,width=\linewidth]{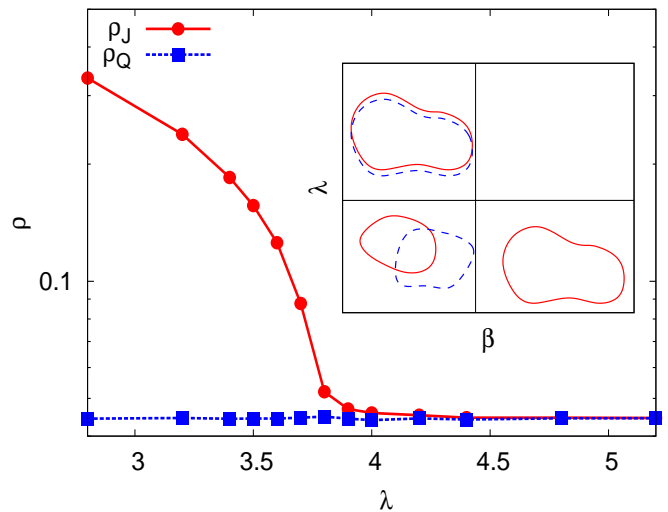}
\caption{Inset: Bulk phase diagram for the model in Eqs.~(\ref{action})~and~(\ref{SHeis}) with bosons and Heisenberg spins. The phase diagram is mathematically equivalent to a system of decoupled currents and spins, hence the straight line boundaries. At $\lambda=0$, the system has a paramagnet-ferromagnet transition as $\beta$ is increased, while the bosons are superfluid throughout. As $\lambda$ is increased, the boson currents bind to the hedgehog currents. The loop pictures in the phases show a ``snapshot'' of the phase. Red solid loops mean that boson currents are proliferated in the phase, while blue dashed loops indicate proliferated hedgehog currents. The phase of interest is the ``binding'' phase in the upper left corner where bosons are bound to hedgehogs. The main figure shows the current-current correlations of the bosons and hedgehogs as $\lambda$ is increased while $\beta=0$, for a system of linear dimension $L=6$. We see that the correlators become essentially equal as the system enters the upper left phase, indicating that bosons have bound to hedgehogs.}
\label{heisbulk}
\end{figure}

The bulk phase diagram of this model is shown in the inset in Fig.~\ref{heisbulk} and can be understood essentially analytically.  Indeed, in the case where the entire system is in the same phase (i.e., the boson-hedgehog binding is applied everywhere), we can change to new variables in the partition sum
\begin{equation}
\tilde J_\mu(r) \equiv J_\mu(r) - Q_\mu(r) ~,
\label{shift}
\end{equation}
which satisfy the same conditions as the original boson currents.  Expressed in these variables, the first and second terms in Eq.~(\ref{action}) decouple, and we can study them independently. 
At small $\beta$, the $\vec{n}$ spins are disordered, which also implies that the hedgehog currents are proliferated. As $\beta$ is increased, the spins order. This means that there is a large energy cost for hedgehog currents to exist. We say that the hedgehog currents are gapped, and only small loops of them are present. We can determine the location of the spin-ordering phase transition by finding singularities in the heat capacity, or by performing finite-size scaling on the magnetization as described above. The value found by our numerics agrees with the literature on the 4D classical Heisenberg model.\cite{McKenzie2}

At small $\lambda$, the $\tilde J_\mu$ variables are proliferated. In our original variables, this means that the boson currents are effectively independent of the hedgehog currents and are condensed. At large $\lambda$, the $\tilde J_\mu$ variables are gapped. The boson currents are bound to the hedgehog currents. The result of this can be seen in Fig.~\ref{heisbulk}, which shows the current-current correlators for the boson and hedgehog currents. For small $\lambda$, there is no relation between the bosons (red solid line) and the hedgehogs (blue dashed line). As $\lambda$ is increased, the current-current correlators become essentially identical as the bosons and hedgehogs are bound together. We can determine the location of the phase transition in $\lambda$ by studying singularities in the heat capacity, or by peforming finite-size scaling on $\rho_{\tilde J}$ as described above. 

We can now summarize the phase diagram shown in the inset in Fig.~\ref{heisbulk}. At small $\lambda$ and $\beta$ both the boson and hedgehog currents are proliferated and are not bound, while at large $\lambda$ and $\beta$ all currents are gapped. At small $\lambda$ and large $\beta$ the boson currents are condensed but the hedgehog currents are gapped. Finally, when $\lambda$ is large and $\beta$ is small the system is in the ``binding'' phase with ``proliferated'' bound states, which we will argue is a (3+1)D topological phase protected by the appropriate symmetries.

It is also helpful to think about the ``easy-plane'' regime for the spin variables, in which the spins, $\vec{n} = (n_a, n_b, n_c)$, are roughly in the $ab$-plane, with only small $c$ components; we will denote the corresponding global symmetry of spin rotations in the $ab$-plane as $U(1)_{\rm spin}$. In the easy-plane case we can define ``vortices'' of the XY spins $(n_a, n_b)$ (i.e., phase windings of the complex order parameter $\sim n_a + i n_b$). 
The vortices are defined on the plackets of the cubic lattice. Therefore in the (3+1)D space-time they are represented as two-dimensional ``world-sheets.'' When dealing with vortices, one can gain intuition by thinking in terms of only the three spatial dimensions of our (3+1) dimensional space-time. In this picture the bosons and hedgehogs are represented by point particles, while the vortices are represented by lines. The next two paragraphs can be most easily understood by thinking in this picture.

\begin{figure}
\includegraphics[width=\linewidth]{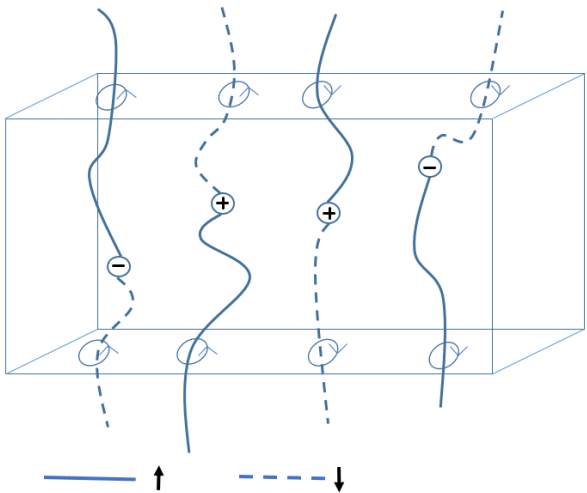}
\caption{As discussed in the text, the easy-plane spin system has two species of vortex lines, $\uparrow$ and $\downarrow$, depending on $n_c$ at the core. A hedgehog is a transition point between the two types of lines, with the sign of the hedgehog determined by the orientation of the type and vorticity of the vortex lines, as shown with examples. This figure shows only the spatial dimensions of the system, therefore the hedgehogs are point particles and the vortices are lines. Applying a Zeeman field to the surface means allowing only one type of vortex line through the surface. This leads to a correlation between the hedgehog number (and therefore the boson charge) and the vorticity at the surface, which is the origin of the Hall conductivity.}
\label{monopoles}
\end{figure}

Though ordinary XY spins are not defined at the core of a vortex, our spins have a $c$-component which can point either up or down at the vortex core. We can define two species of vortices, which we call the $\uparrow$ and $\downarrow$ species, depending on whether $n_c$ is positive or negative at the core. This description is useful since an $\uparrow$ vortex ending and continuing as a $\downarrow$ vortex is a hedgehog. Therefore our system can be thought of as a system of vortex lines having two different species. These vortex lines can change species, and the locations where this happens are hedgehogs.\cite{LesikSenthil} This is illustrated in Fig.~\ref{monopoles}.

Hedgehog number can be either positive or negative, and this is determined by the properties of the vortex line the hedgehog is attached to.
If, when looking along the line from the $\uparrow$ core to the $\downarrow$ core the vorticity is clockwise (counterclockwise), we define a positive (negative) hedgehog.  This is pictured in Fig.~\ref{monopoles}.

\subsubsection{Importance of Discrete Symmetry}
\label{subsubsec:HeisSym}

The action in Eq.~(\ref{action}) has a $U(1)$ symmetry which comes from the conservation of the $J_\mu$ currents; this is boson charge conservation symmetry. It also has an $SO(3)$ symmetry from the spins. Both of these symmetries participate in the protection of the topological phase, in the sense that if they are broken it is possible to continuously connect the topological phase to a trivial phase.
In addition, the action has a $\ztwo$ symmetry, which is obtained by reflecting the $\vec{n}$ spins in a plane in the spin space. To see how this affects the hedgehog current, we can examine Eq.~(\ref{alpha}), taking the reference vector $\vec{N}_0$ to be in the plane of reflection. We see that reflecting the spins changes the sign of the imaginary part of $e^{i\alpha_\mu}$, and therefore the hedgehog current changes sign under such a reflection. For our entire action to be invariant, we therefore need to combine such reflections with an operation which changes the sign of the boson currents. For concreteness we will consider the $\ztwo$ symmetry corresponding to reflections of the $\vec{n}$ variables in the $ab$ plane of the spin space. In this case the $\ztwo$ symmetry can be summarized as:
\begin{equation}
\begin{array}{ccc}
n_a,n_b & \rightarrow & n_a,n_b \\
n_c & \rightarrow & -n_c\\
Q_\mu & \rightarrow & -Q_\mu\\
J_\mu & \rightarrow & -J_\mu 
\end{array}.
\label{ztwoeqn}
\end{equation}
Note that it is also possible to reflect the $\vec n$ spins around a different plane, but this is not a distinct symmetry since it is simply the product of the above $\ztwo$ symmetry and an element of $SO(3)$. 

By analogy with the electronic topological insulator, we would like the $\ztwo$ symmetry described above to be a ``time-reversal'' symmetry, i.e.~it should be anti-unitary. The symmetry in Eq.~(\ref{ztwoeqn}) can be either a unitary or anti-unitary symmetry. Note that Eq.~(\ref{action}) is a real, (3+1)-dimensional action which is assumed to arise from the Trotter decomposition of the imaginary time propagator (i.e., Euclidean path integral) of a three-dimensional quantum Hamiltonian. The symmetry operations in Eq.~(\ref{ztwoeqn}) can be derived from the action of a symmetry operation on the quantum Hamiltonian. Therefore asking whether the symmetry in Eq.~(\ref{ztwoeqn}) is anti-unitary is the same as asking whether the symmetry of the quantum Hamiltonian which generates Eq.~(\ref{ztwoeqn}) is anti-unitary. This is a difficult question for us to answer as we do not know the quantum Hamiltonian which has Eq.~(\ref{action}) as its Euclidean path integral.  Nevertheless, we take the perspective where we can check whether the original Hamiltonian has time reversal by complex-conjugating the action combined with the appropriate variable transformations, and in this way we can view the above symmetry of the action also as time reversal.

In the easy plane case, our system has $U(1)_{\rm boson}\times U(1)_{\rm spin}$ in addition to this discrete symmetry. We can think of the $U(1)_{\rm spin}$ as also coming from a boson, and $n_a+i n_b$ gives the phase degree of freedom of this boson. We can imagine allowing tunnelling between the two $U(1)$ symmetries. If we do this, then the discrete symmetry should act the same way on the two species, and we can see that the above symmetry changes the number of the bosons but not their phase. In a system consisting of one species of bosons, the symmetry which acts in this way is anti-unitary.
From now on the symmetry described above will also be treated as anti-unitary and denoted by $\ztwot$.

The action in Eq.~(\ref{action}) is invariant under several $\ztwo$ symmetries, but for our purposes we will consider only the $\ztwot$ symmetry described above as important, as it protects the topological behavior. To see this we can break this symmetry and argue that the topological phase is destroyed.
We break the $\ztwot$ symmetry by introducing a Zeeman field into our action:
\begin{equation}
S_{\rm Zeeman}=-h\sum_R n_c(R).
\label{Zeeman}
\end{equation}
Here $h$ is the strength of the Zeeman field, which points in the $c$-direction. Note that the Zeeman field does not break $U(1)_{\rm spin}$ or $U(1)_{\rm boson}$. In our picture of two species of vortices (Fig.~\ref{monopoles}), the Zeeman field forbids one of the species. Since the vortex lines cannot change species, hedgehogs are forbidden and the binding phase is destroyed. 
This can be made more precise if we replace the binding term in Eq.~(\ref{action}) with the following term:
\begin{equation}
\frac{\lambda}{2}\sum_{r,\mu} [ J_\mu(r) - \eta Q_\mu(r)]^2 ~,
\label{tbind}
\end{equation}
where $\eta$ is a real number. If we choose the parameters $\beta$ and $\lambda$ so that the system is initially in the binding phase, the introduction of $\eta$ allows us to tune the system between the binding phase ($\eta=1$) and the trivial insulator ($\eta=0$). Without a Zeeman field, the system undergoes a phase transition as $\eta$ is changed between $1$ and $0$. 
When a Zeeman field is applied, the hedgehogs are effectively forbidden, and so we can tune parameters $h$ and $\eta$ without going through a phase transition. Indeed, when $\eta=1$, the change of variables in Eq.~(\ref{shift}) leads to decoupled $\tilde{J}$ currents and spins, so there is no phase transition when making $h$ arbitrarily large to align all spins. We can then tune $\eta$ to zero and finally reduce $h$ back to zero, all without undergoing a phase transition.  Thus, in the presence of the Zeeman field, the phase with $\eta=1$ is not distinct from the trivial insulator.

\subsubsection{Binding of Multiple Bosons to a Hedgehog}
The above methods also allow us to answer the question of what happens to the system if $\eta$ in Eq.~(\ref{tbind}) is an integer larger than 1.  In a $U(1)\times U(1)$ system in two dimensions, we studied the binding of multiple bosons to vortices (realized by taking integer $\eta$) and found that each number of bound bosons led to a different symmetry-protected topological phase.\cite{FQHE} There were therefore as many SPTs as there are integers, in agreement with the cohomology classification.\cite{WenScience,*WenPRB, LuVishwanath} In the present three-dimensional case the classification of Chen et al.\cite{WenScience,*WenPRB} for $U(1)$ and $\ztwot$ symmetry predicts the existence of only a finite number of such symmetry protected topological phases, implying that not every value of $\eta$ would lead to a distinct phase. Indeed, we find that all systems with $\eta$ an even integer are topologically trivial, while when $\eta$ is odd we have the same topological phase as $\eta=1$. 

We can justify this claim by showing that $\eta=2$ can be continuously connected to $\eta=0$ which is a trivial insulator. This argument can then be extended to show that any two systems where $\eta$ differs by $2$ are in the same phase.  Our argument is inspired by Ref.~\onlinecite{BiRasmussenXu13}, except that here we are working with a microscopic model rather than a topological field theory.
We start by considering two copies of our action, each with its own bosons and Heisenberg spins and with $\eta=1$:
\begin{eqnarray}
&&S=-\beta\sum_{R,\mu}\left[ \vec{n}^{(1)}(R)\cdot \vec{n}^{(1)}(R+\hat{\mu})+\vec{n}^{(2)}(R)\cdot \vec{n}^{(2)}(R+\hat{\mu})\right]\nonumber\\
&&+\frac{\lambda}{2}\sum_{r,\mu}\left( [ J_\mu^{(1)}(r)- Q_\mu^{(1)}(r)]^2+[ J_\mu^{(2)}(r)- Q_\mu^{(2)}(r)]^2\right),
\label{doubleaction}
\end{eqnarray}
where the superscripts indicate which copy a variable is from.  We now couple the two copies by adding the following terms:
\begin{eqnarray}
\delta S&=&-A\sum_{R} \vec{n}^{(1)}(R)\cdot \vec{n}^{(2)}(R)\nonumber\\
&-&B\sum_{r} \cos[\Phi^{(1)}(r)-\Phi^{(2)}(r)].
\label{AB}
\end{eqnarray} 
When $A$ is large and positive, the first term above locks spins of the different copies together, $\vec{n}^{(1)}\approx\vec{n}^{(2)}$. The hedgehog variables therefore take on the same values, and either can be viewed as the hedgehog number of the whole spin system in this case: $Q \approx Q^{(1)} \approx Q^{(2)}$. On the other hand, when $A$ is large and negative the spins of different types are locked in opposite directions,  $\vec{n}^{(1)}\approx -\vec{n}^{(2)}$, and $Q^{(1)} \approx -Q^{(2)}$.

The $\Phi$ variables are $2\pi$-periodic phases that can be thought of as conjugates to the $J_\mu$ variables. More precisely, in our path integral we sum over only the configurations of $J_\mu$ in which the currents are divergenceless. We can instead sum over all configurations of $J_\mu$, and include the following term in our path integral:
\begin{equation}
\int_0^{2\pi} D\Phi(r) e^{-i\sum_r \Phi(r)(\sum_\mu\nabla_\mu J_\mu)(r)} ~,
\end{equation}
which dynamically enforces the constraint that the boson currents be conserved. We introduce $\Phi$ variables for each copy of the boson currents, and the $B$ term is tunnelling between the two copies. 
When $B$ is large (of either sign) only the sum of $J^{(1)}$ and $J^{(2)}$ is conserved and can be identified as the current of the whole boson system (i.e., combining both copies).

When $A$ and $B$ are large and positive, we can expand the terms on the second line of Eq.~(\ref{doubleaction}), and take $J = J^{(1)} + J^{(2)}$, $Q = Q^{(1)} = Q^{(2)}$. In this case we obtain coupling between $J$ and $Q$ which is effectively the same as in Eq.~(\ref{tbind}) with $\eta=2$. 
On the other hand, when $A$ is large and negative, but $B$ is still large and positive, $J$ is unchanged but its coupling to the hedgehogs vanishes as contributions from $Q^{(1)}$ and $Q^{(2)}$ cancel; this gives Eq.~(\ref{tbind}) with $\eta=0$.
We can continuously deform $A=0$, $B=0$ to $A=\infty$, $B=\infty$ without undergoing a phase transition. In addition, we can deform $A=0$, $B=0$ to $A=-\infty$, $B=\infty$ without undergoing a phase transition. This implies that we can tune from $\eta=2$ to $\eta=0$ without undergoing a phase transition, and so both of these cases are in the trivial insulating phase.

\subsection{Phase Diagram on the Boundary Between the Binding Phase and a Trivial Insulator}
\label{subsec:heissurf}
By analogy to the fermionic topological insulator, we expect that one way to investigate the topological nature of our phase is to study the physics of its surface. In particular, a number of interesting phases have been predicted on the surface of the bosonic TI,\cite{SenthilVishwanath} and if we can identify these phases on the surface of our binding phase, it will be a powerful argument that the binding phase is a bosonic TI.

We introduce a surface between the binding phase and a trivial insulator by allowing $\eta$ in Eq.~(\ref{tbind}) to vary spatially.
We vary $\eta$ in the $z$ direction, so that:
\begin{equation}
\eta[r=(x,y,z,\tau)]=
\left\{ \begin{array}{cc}
1, &~ z_L\leq z<z_R\\
0, &~ \rm{otherwise}\\
\end{array}\right..
\label{eta_r}
\end{equation}
This leads to the binding phase in the region $z_L < z < z_R$, while the trivial phase occupies the rest of the space. Note that there are two surfaces of the binding phase: one at $z_L$ and one at $z_R$. The above geometry in our Monte Carlo setup with periodic boundary conditions corresponds to a four-dimensional torus which is divided into two parts along the $z$-direction.

On the surface, we can measure all of the quantities which we measured in the bulk, but we now only sum over the sites near the surface (and when averaging over directions we only use the $\hat{x}$, $\hat{y}$ and $\hat{\tau}$ directions). 

\begin{figure}
\includegraphics[width=\linewidth]{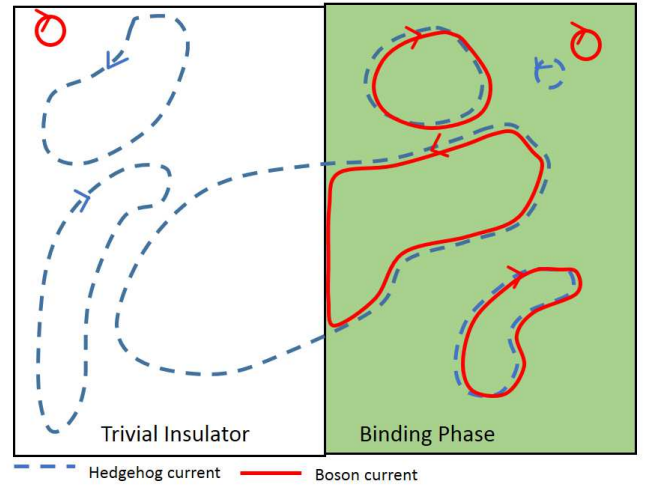}
\caption{A snapshot of the system, when it is spatially divided into a region which is a trivial insulator and a region which is in the binding phase. In the trivial phase, boson currents are gapped and hedgehog currents are proliferated. In the binding phase both currents are gapped individually, but their bound states are proliferated. Large hedgehog current loops can exist in either region; however, when such a loop tries to cross the boundary the conservation of boson current leads to a conflict between the two sides, and this can lead to exotic surface physics. }
\label{surface}
\end{figure}

The binding between hedgehogs and bosons in the bulk of the topological phase leads to exotic physics on the surface. In both the binding and trivial phases, hedgehog currents are proliferated, while boson currents are bound to the hedgehog currents in the binding phase and are absent in the trivial phase, as shown in Fig.~\ref{surface}. Consider what happens when a hedgehog loop tries to cross the boundary between the phases. Since the boson currents must form closed loops, the above conditions cannot both be satisfied without something interesting happening on the surface. For example, we could have unbound boson currents on the surface, or hedgehogs could be effectively forbidden from crossing the surface.

\begin{figure}
\includegraphics[angle=-90,width=\linewidth]{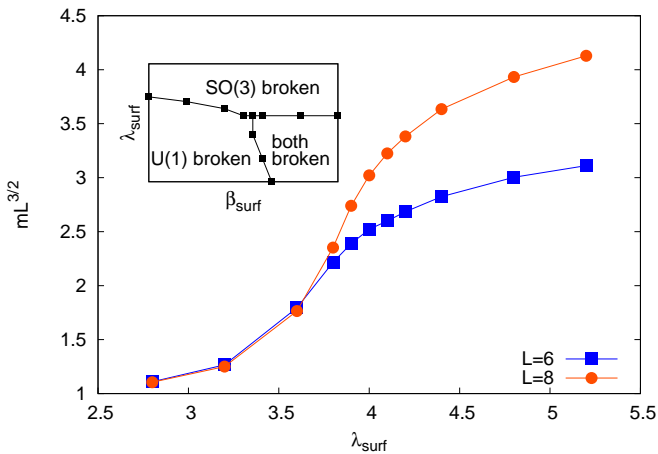}
\caption{The inset shows the phase diagram of the surface of the binding phase, without a Zeeman field. It was obtained by tuning the bulk parameters deep into the binding phase and then varying $\beta$ and $\lambda$ {\em only on the surface}.  We find that in this model our surface always spontaneously breaks a symmetry. At small $\beta_{\rm surf}$ and $\lambda_{\rm surf}$ the boson $U(1)$ symmetry is broken and the bosons condense into a superfluid, while at large $\lambda_{\rm surf}$ the spin $SO(3)$ symmetry is broken and the spins align into a ferromagnet. At small $\lambda_{\rm surf}$ and large $\beta_{\rm surf}$ both symmetries are broken.  The main plot shows surface magnetization on a sweep in $\lambda_{\rm surf}$ for $\beta_{\rm surf}=0$. We show $mL^{3/2}$, which is independent of system size in the disordered phase, and grows with system size in the ordered phase. We can clearly see that the $SO(3)$ symmetry is broken as $\lambda_{\rm surf}$ is increased past a value of approximately $4$. All data in this section was taken with $\beta_{\rm bulk}=0$, $\lambda_{\rm bulk}=5.2$.}
\label{heissurf}
\end{figure}

In order to study the surface physics and search for the exotic surface behavior predicted in Ref.~\onlinecite{SenthilVishwanath}, we first determine the surface phase diagram.
To do this, we fix the values of $\beta$ and $\lambda$ in the bulk and tune them only on one of the surfaces, by setting:
\begin{eqnarray}
\beta_\mu(X,Y,Z,T) &=&
\left\{
\begin{array}{ll}
\beta_{\rm surf}, & Z = z_R - 1/2, ~\mu=\hat{x},\hat{y},\hat{\tau}\\
\beta_{\rm bulk}, & {\rm otherwise}
\end{array}\right. ~
\label{bulkvsurf} \\
\lambda_\mu(x,y,z,\tau) &=&
\left\{
\begin{array}{ll}
\lambda_{\rm surf}, & z = z_R, ~\mu=\hat{x},\hat{y},\hat{\tau}\\
\lambda_{\rm bulk}, & {\rm otherwise }
\end{array}\right. ~.
\end{eqnarray}
Here we focused on the surface at $z_R$, and $\mu$ denotes link orientation for either $\beta$ or $\lambda$ terms.  We show the surface phase diagram in the inset of Fig.~\ref{heissurf}.  All data was taken with $\beta_{\rm bulk}=0$, $\lambda_{\rm bulk}=5.2$, parameters which put the bulk deep into the binding phase.  The surface phase diagram contains three distinct phases. When $\lambda_{\rm surf}$ is small the bosons are in a superfluid phase, breaking their $U(1)$ symmetry. This is the scenario pictured in Fig.~\ref{surface}, where hedgehogs can cross the surface, and these crossings are connected by boson currents. If $\beta_{\rm surf}$ is also small the $SO(3)$ symmetry is unbroken as the spins are disordered. As $\beta_{\rm surf}$ increases at small $\lambda_{\rm surf}$, the $SO(3)$ symmetry breaks, and so both symmetries are broken. Finally, at large $\lambda_{\rm surf}$ bosons see a large energy cost, and so the $U(1)$ symmetry is unbroken. This forbids hedgehogs from crossing the surface, leading to a system of $SO(3)$ spins with hedgehogs forbidden, which on the three-dimensional cubic latice is known to have magnetic order thus breaking the $SO(3)$ symmetry.\cite{LauDasgupta, LesikAshvin}

The locations of the phases and phase transitions in Fig.~\ref{heissurf} were determined by studying singularities in the specific heat, as well as by studying the surface magnetization and current-current correlators. As an example of such data, in the main plot of Fig.~\ref{heissurf} we show the magnetization, multiplied by the square-root of the volume of the surface ($L^{3/2}$). This quantity should be constant when the spins are disordered and should grow with system size when they are ordered. The data was taken with $\beta_{\rm surf}=0$ and increasing $\lambda_{\rm surf}$. We can see that the spins order at $\lambda_{\rm surf} \approx 4$. 

\begin{figure}
\includegraphics[angle=-90,width=\linewidth]{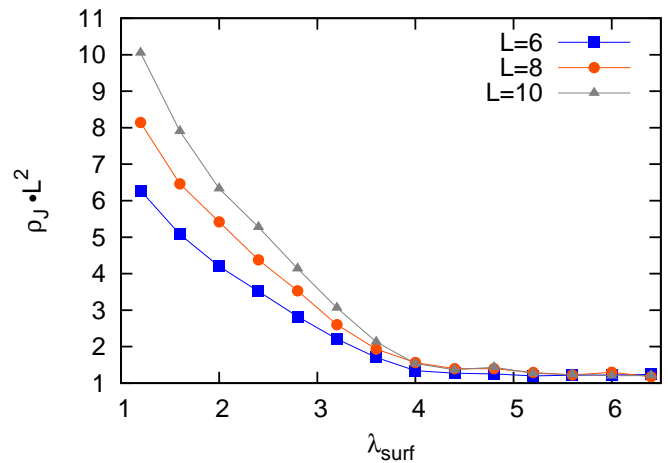}
\caption{Current-current correlators of the entire system, multiplied by $L^2$, used to detect symmetry breaking of the bosons on the surface. This quantity is constant when the boson $U(1)$ symmetry is preserved and increases linearly with system size when it is broken on the surface. Parameters used are the same as in Fig.~\ref{heissurf}. We see that there is a phase transition at $\lambda_{\rm surf}\approx 4$.
}
\label{slabcurs}
\end{figure}

In order to use the current-current correlators to detect the breaking of the $U(1)$ boson symmetry on the surface, we must think through such measurements carefully.  In a $D$-dimensional system of boson worldlines (space-time currents), the argument that $\rho_J(k_{\rm min}) \sim k_{\rm min}^2 \sim 1/L^2$ in the gapped phase relies on the conservation of the currents $J$.  However, this is no longer true if we consider only the surface, as currents can enter and exit from the rest of the system.  
Therefore we instead measure the current-current correlators in the entire system (though only in the $\hat{x}$, $\hat{y}$, and $\hat{\tau}$ directions, which are parallel to the surface). When currents are gapped everywhere, this quantity will still be proportional to $1/L^2$. 
If we measured the current-current correlator only on the surface in the phase where the surface is a superfluid, we would expect the result to be independent of system size. Since we are measuring the correlator over the whole system, the result should be proportional to the fraction of the system which makes up the surface, which is $1/L$.
In Fig.~\ref{slabcurs} we plot $\rho_J \cdot L^2$. We see that at large $\lambda_{\rm surf}$ this quantity is independent of system size, which tells us that the currents are gapped everywhere and the surface is an insulator in the boson degrees of freedom. At small $\lambda_{\rm surf}$, $\rho_J\cdot L^2 \sim L$, which tells us that there is a region whose volume fraction is proportional to $1/L$ where the bosons are in a superfluid phase. We interpret this as evidence that there is a superfluid at the surface.

The surface phase where the $U(1)$ boson symmetry is broken is clearly a superfluid of bosons. In the easy-plane picture the spins can also be thought of as having a $U(1)$ symmetry which is broken when the spins are ordered, and so the phase where the spins order can also be thought of as a superfluid of new particles representing the spins. We can now ask whether these superfluids are trivial superfluids, or the exotic superfluids thought to exist on the surface of a bosonic TI.\cite{SenthilVishwanath} The exotic properties have to do with the charge and statistics of gapped vortex excitations, and unfortunately we do not have access to these properties in our Monte Carlo. Therefore our surface superfluids cannot tell us whether the binding phase is a bosonic TI.

One predicted feature of the superfluids on the surface of the bosonic TI is that they are connected by a direct transition which is a deconfined critical point.\cite{SenthilVishwanath} As shown in Figs.~\ref{heissurf} and \ref{slabcurs}, our $SO(3)$ and $U(1)$ symmetries appear to break on the opposite sides of the same point, and so it looks that we also have a direct transition between these phases; however, we would need larger system sizes to see whether there is indeed a direct transition and whether it is continuous.
Though this does not definitively show that the binding phase is topological, but it does provide evidence that we have an unusual field theory on the surface.

\subsection{Surface with Zeeman Field}
Another exotic phase predicted to exist on the surface of the bosonic TI is a phase which breaks $\ztwot$ symmetry and has a quantized Hall reponse. We can try to realize this phase by applying a Zeeman field on the surface of our model to explicitly break the $\ztwot$ symmetry. We add a term similar to Eq.~(\ref{Zeeman}) to our action, but only on the surface of the model. We then expect that there is a surface phase which does not break any $U(1)$ symmetry---i.e., an insulator---but which has the surface Hall conductivity quantized to an odd integer, which is different from the even values expected in the (2+1)D bosonic integer quantum Hall effect. For easy-plane spins, we expect that a finite Zeeman field is needed to destroy the surface superfluids to reach this phase, while for SO(3) spins used here and starting in the spin-ordered phase, we expect to immediately transition to the quantum Hall insulator.\cite{LesikAshvin}

In our Monte Carlo study, we actually apply Zeeman field $h$ to both surfaces but take the fields on the two surfaces to have opposite signs.  We take $\lambda=5.2$ and $\beta=0$ everywhere (including at the surfaces), which puts the bulk regions deep into the binding or trivial phases respectively, while the surfaces start in the spin-ordered phase at $h=0$.

Since the Hall conductivity in the surface system of spins and bosons will be due to correlations between the vorticity of the spins and the boson charges,\cite{FQHE} we will measure these correlations directly before moving on to the more complicated Hall conductivity measurement. 
In the absence of the Zeeman field, we can use the $\ztwot$ symmetry, which is reflection of the $\vec n$ variables in the $ab$-plane, to change the sign of hedgehog and hence the boson charge without changing the spin vorticity, and so such correlations vanish. This can be seen in Fig.~\ref{monopoles}, where, for example, on the top surface there is one clockwise vortex attached to a positive hedgehog (which is in turn bound to a positively charged boson), and another clockwise vortex is attached to a negative hedgehog. Applying a Zeeman field corresponds to only allowing one type of vortex ($\uparrow$ or $\downarrow$) to pass through the surface. In Fig.~\ref{monopoles}, this means that only solid lines are allowed to pass through the top surface. We can see that this leads to a correlation between the vorticity of the vortex on the surface and the charge of the boson it is binding nearby. 

We can further think about Fig.~\ref{monopoles} as depicting a slab of the binding phase. Opposite Zeeman fields on the two surfaces give us a quasi-two-dimensional slab on which vortices are bound to charges with definite relation between the vorticity and charge, and these bound states are proliferated. We have studied such a system in a previous work,\cite{FQHE} and found that its Hall conductivity is quantized to be equal to two. It is reasonable to assume that this conductivity is evenly distributed between the two surfaces, leading to the surface Hall conductivity equal to one on each surface. This intuitive argument reproduces the prediction of Ref.~\onlinecite{SenthilVishwanath}.

\begin{figure}
\includegraphics[width=\linewidth]{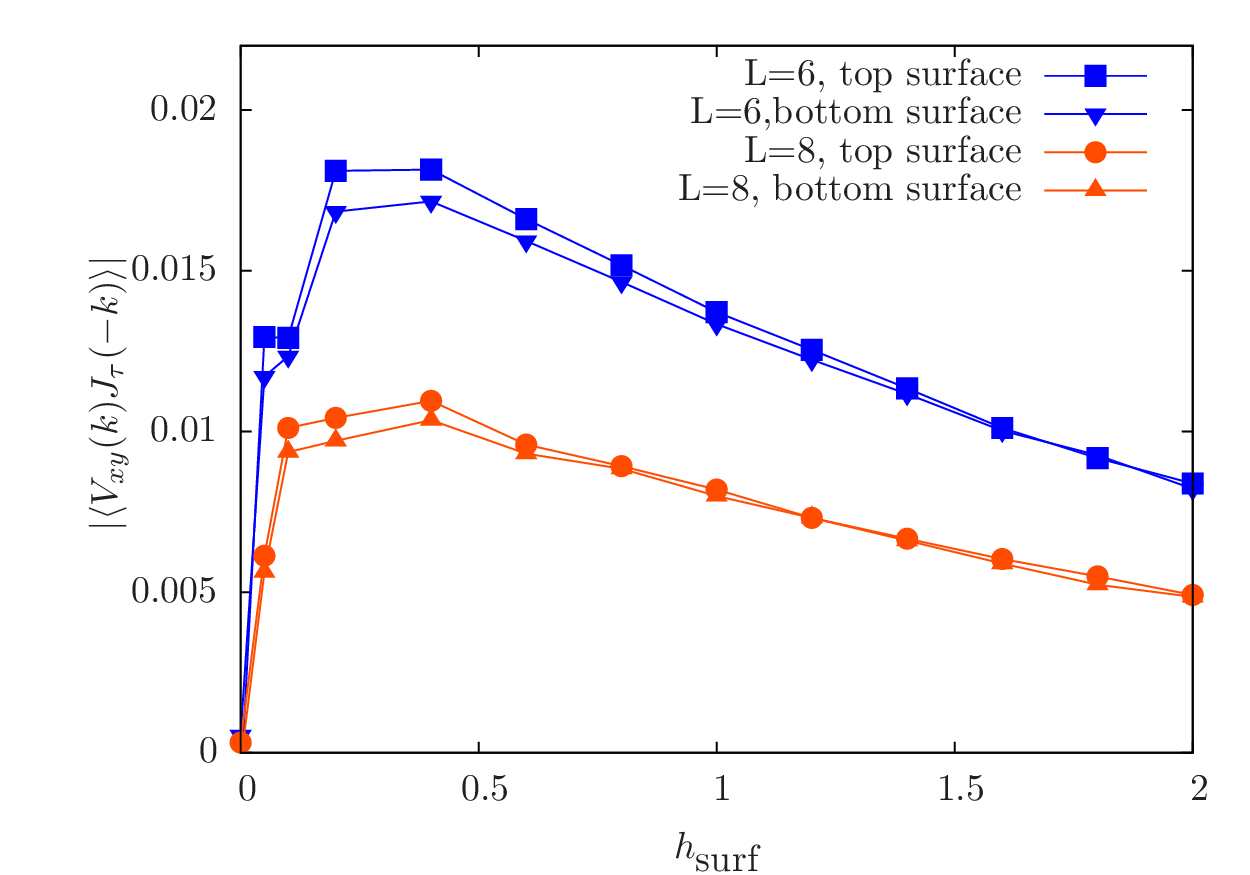}
\caption{Spin vortex - boson charge correlators near the surface of the binding phase in the model Eqs.~(\ref{action})~and~(\ref{SHeis}) using Heisenberg spins. The horizontal axis is the strength of the Zeeman field applied only near the surfaces (and of opposite sign on the two surfaces). We see that as soon as the Zeeman field is turned on, the correlator takes a non-zero value. The value is non-universal, but is approximately the same on the top and bottom surfaces. Data was taken with $\beta=0$, $\lambda=5.2$ everywhere and the binding phase occupying half of the system, cf.\ Eq.~(\ref{eta_r}).  Since we do not know how to properly couple external gauge fields to this system, we are unable to compute the Hall conductivity.
}
\label{heishall}
\end{figure}

To measure correlations between vortices and bosons we first define the spin vorticity $V_{\mu\nu}(R)$ as
\begin{equation}
V_{\mu\nu}(R) = \frac{1}{2\pi}[\nabla_\mu s_{\nu}(R) - \nabla_\nu s_{\mu}(R)] ~,
\label{Vmn}
\end{equation}
where $s_\mu(R) \in (-\pi, \pi]$ measures the difference between the spin angles at $R+\hat{\mu}$ and $R$ and brings it to $(-\pi,\pi]$:
\begin{equation}
s_{\mu}(R)\equiv \left[\tan^{-1}\left(\frac{n_b(R+\hat{\mu})}{n_a(R+\hat{\mu})}\right) -\tan^{-1}\left(\frac{n_b(R)}{n_a(R)}\right)\right]{\rm mod}~2\pi.\nonumber
\end{equation}
We can then Fourier transform the vorticity as follows:
\begin{equation}
V_{xy}(k) = \frac{1}{\sqrt{L^3}}{\sum_{R}}^\prime V_{xy}(R) e^{-ik\cdot R}.
\end{equation}
The prime on the sum indicates we are summing over all sites at a fixed $z=z_R$. We measure $\left|\la V_{xy}(k_{\rm min})J_\tau(-k_{\rm min})\ra\right|$, where $k_{\rm min}=(2\pi/L,0,0,0)$, and the results are shown in Fig.~\ref{heishall}. 

We see that as soon as the Zeeman field is applied, the vortices and charges become correlated. Unlike the Hall conductivity, we do not expect these correlations to approach any universal value. We do note that the correlations on the two surfaces of the system are approximately equal, which is encouraging as we would expect the Hall conductivity on these surfaces to be equal as well. The differences between the correlations on the two surfaces is likely due to the fact that the surfaces are realized differently on a lattice, but we would expect that universal values such as the Hall conductivity would not have these differences.

In order to measure Hall conductivity, we need to couple both the bosons and spins to external probing gauge fields, and then use linear response theory to determine the conductivity. When we do this we run into a problem, which has to do with the way the hedgehog currents $Q_\mu$ were defined. We can see from Eq.~(\ref{omega}) that $\omega_{\mu\nu}$ is a discontinuous function of spins since we required it to be brought to $(-\pi, \pi]$. When including the probing gauge fields, this causes the path integral to be a discontinuous function of the probing fields, which prevents us from taking the derivatives needed for linear response theory, and so we do not know how to calculate conductivity in this system. In the next section we will find a way around this problem, while in the present case we can only appeal to the intuitive argument presented above.

\section{Realizing the topological insulator by binding bosons to hedgehogs of an easy-plane \cp model}
\label{section::CP1}

The Heisenberg model discussed in the previous section allowed us to realize a ``binding'' phase of bosons and hedgehogs. However, in this model we were unable to find definitive evidence that the binding phase is in fact a symmetry-protected bosonic topological insulator, although our indirect arguments are compelling. In this section we introduce a new model, which includes the spin degrees of freedom in a \cp representation. 
This theory has spinor matter fields (``spinons'') coupled to a compact gauge field and is a faithful representation of the microscopic spin system with short-range interactions (i.e., such a lattice ``field theory'' is ``emergable'' from a local microscopic Hamiltonian). We will see that this formulation allows us to make measurements, such as a Witten effect in the bulk and a quantized Hall conductivity on the surface, which indicate that the binding phase is a bosonic TI.

\subsection{Bulk Phase Diagram}
The following action represents the spins in the $CP^1$ representation:
\begin{eqnarray}
&&S_{\rm spin}=-\beta\sum_{s=\uparrow,\downarrow}\sum_{R,\mu} [z_s^\dagger(R)z_s(R+\hat\mu)e^{-ia_\mu(R)}+c.c.] \nonumber\\
&&-K\sum_{R,\mu<\nu} \cos[\nabla_\mu a_\nu(R)-\nabla_\nu a_\mu(R)].
\label{cp1action}
\end{eqnarray} 
Here the spins are represented by two complex bosonic fields $z_\uparrow$,$z_\downarrow$ (``spinons''), which satisfy $|z_\uparrow|^2+|z_\downarrow|^2=1$. We can write the $z$ fields as a spinor, $\mathbf{z}\equiv(z_\uparrow,z_\downarrow)^T$, and extract the spin $\vec{n}=\mathbf{z^\dagger} \vec\sigma \mathbf{z}$, where $\vec{\sigma}\equiv (\sigma_1,\sigma_2,\sigma_3)$ is a vector of Pauli matrices.
The spinon fields are minimally coupled to a compact gauge field $a_\mu(R)$. The last term is a Maxwell-like term for the compact gauge field, which appears after partially integrating out the spinon fields. The variables in the above action live on a cubic lattice, where $R$ gives the position on the lattice and $\mu$,~$\nu$ are directions.

The \cp model defined above actually has global $SO(3)$ symmetry, similar to the previous section. In this section we find it convenient to break the $SO(3)$ symmetry down to $U(1)$ explicitly by taking the ``easy-plane'' limit of the $CP^1$ model. We align all the spins $\vec{n}$ in the $ab$-plane, which corresponds to fixing the magnitude of $z_\uparrow$ and $z_\downarrow$, and allowing only phase fluctuations, i.e., $z_s\equiv \frac{1}{\sqrt{2}}e^{i\phi_s}$. The $CP^1$ model in the easy-plane limit becomes:
\begin{eqnarray}
&&S_{\rm spin}=-\beta\sum_{s=\uparrow,\downarrow}\sum_{R,\mu} \cos[\nabla_\mu\phi_s(R)-a_\mu(R)]\nonumber\\
&&+\frac{K}{2}\sum_{R,\mu<\nu}\left[\nabla_\mu a_\nu(R)-\nabla_\nu a_\mu(R)-2\pi B_{\mu\nu}(R)\right]^2.
\label{sspin}
\end{eqnarray}
Here $\phi_\uparrow$ and $\phi_\downarrow$ are $2\pi$-periodic variables which represent the phases of the spinon fields. We have also replaced the cosine in the Maxwell term by a quadratic ``Villain'' form, with $B_{\mu\nu}$ which are unconstrained, integer-valued dynamical variables residing on the plackets of the lattice. Upon summing over $B_{\mu\nu}$ in the partition sum, the third term generates a $2\pi$ periodic function of $\nabla_\mu a_{\nu}-\nabla_\nu a_{\mu}$, and therefore this does not change the universality class of the problem.

Using the Villain form of the Maxwell term is advantageous as it allows us to define the hedgehog current:
\begin{equation}
Q_\mu(r)=\frac{1}{2}\epsilon_{\mu\nu\rho\sigma}\nabla_\nu B_{\rho\sigma}.
\label{mondef}
\end{equation}
Note that $Q_\mu(r)$ resides on the links of the lattice whose sites are labelled by $r$ which, as in the previous section, is interpenetrating with the lattice labelled by $R$. The above definition is analogous to Eq.~(\ref{monopoledef}) with $B_{\rho\sigma} \leftrightarrow \omega_{\rho\sigma}/(2\pi)$. The $B_{\mu\nu}$ variables have the meaning of ``Dirac strings'' of the hedgehogs. Thus defined, the hedgehogs in the \cp model are actually monopoles of the compact gauge field $a_\mu$. We will continue to call them hedgehogs in this work to avoid confusion with a different type of monopole introduced later. 

We can again study the model described by Eqs.~(\ref{action}) and (\ref{sspin}) in Monte Carlo. Equilibration becomes difficult in the regime where $K$ and $\lambda$ are large, and it is necessary to include composite updates which simultaneously change multiple variables. One such update is to change $B_{\mu\nu}$ while also changing $J_\mu$ so that there is no change in the second term in Eq.~(\ref{action}). Another update is to change both $a_\mu$ and $B_{\mu\nu}$ in such a way as to keep the $K$ term in Eq.~(\ref{sspin}) small. 

We can find phase transitions in this model by looking for singularities in the specific heat, which is defined in the same way as in the previous section. We can identify order in the bosonic degrees of freedom by studying the superfluid stiffness and order in the spins by studying the magnetization. In this easy-plane version of the model, the spin degree of freedom is an $XY$ vector with components $(n_a,n_b)$.  Since $(n_a, n_b) = (\cos\phi_{\rm spin}, \sin\phi_{\rm spin})$ with $\phi_{\rm spin} = \phi_\uparrow - \phi_\downarrow$, the magnetization is given by:
\begin{equation}
m = \frac{\left\langle\left|\sum_R e^{i [\phi_\uparrow(R) - \phi_\downarrow(R)]}\right|\right\rangle}{\rm Vol}. 
\label{cp1mag}
\end{equation}

\begin{figure}
\includegraphics[angle=-90,width=\linewidth]{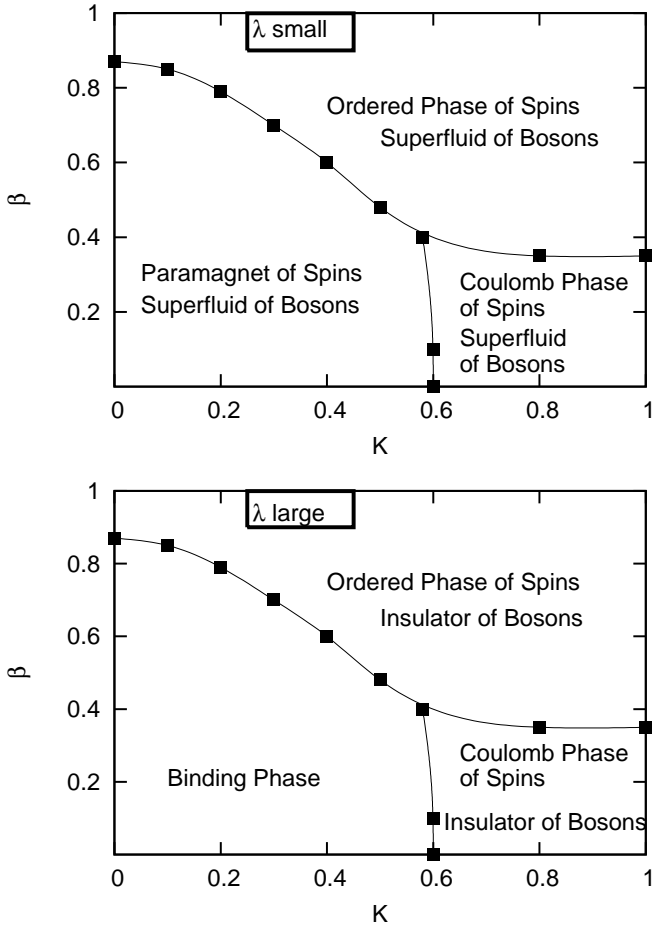}
\caption{Bulk phase diagram in the $\beta$ and $K$ variables for the model defined in Eqs.~(\ref{action}) and (\ref{sspin}) where the spins are described by an easy-plane $CP^1$ model. The top panel shows the phases at small $\lambda$, while the bottom panel shows large $\lambda$. The locations of the phase boundaries are independent of $\lambda$, but the nature of the phases are not. Symbols indicate points where the phase transitions were identified numerically from singularities in the heat capacity. The candidate for the topological phase is the binding phase, which occurs at large $\lambda$, small $\beta$, and small $K$. }
\label{cp1bulkphase}
\end{figure}

The phase diagram in this model is parameterized by $\beta$, $K$, and $\lambda$. As in the previous section, we can make the change of variables in Eq.~(\ref{shift}) and find that the $\tilde{J}$ part of the problem decouples from the spin part. The behaviour of the bosons is the same as in the previous section. When $\lambda$ is small, the physical bosons are essentially independent of the hedgehogs, and are in the superfluid phase. As $\lambda$ is increased, they become bound to hedgehogs. The transition happens at $\lambda\approx 4$. 
The locations of the phase transitions in the spin degrees of freedom are independent of $\lambda$, though the nature of the various phases are not. In Fig.~\ref{cp1bulkphase} we show the phase diagram in the $\beta$ and $K$ variables, for two cases: $\lambda$ small and $\lambda$ large. The phase diagram is consistent with the easy-plane $CP^1$ model in the literature.\cite{artphoton} 

Let us first consider the case when $\lambda$ is small. The bosons will be in a superfluid phase for any $\beta$ and $K$. The spin system has the following three phases: 
{\it i)} When $\beta$ and $K$ are small, the spin degrees of freedom are disordered and the hedgehogs are proliferated. The phase is therefore a conventional paramagnet in the spin degrees of freedom. 
{\it ii)} As $K$ is increased, hedgehogs acquire a large energy cost, and become gapped. This phase was studied in Refs.~\onlinecite{LesikSenthil, LesikAshvin, artphoton}. It is known as the Coulomb phase\cite{HermeleFisherBalents} because it has an emergent gapless photon and gapped excitations that carry charge $1/2$ and interact via a Coulomb interaction. 
{\it iii)} Finally, the phase at large $\beta$ has a large energy penalty for spin fluctuations, and so the spins order. This phase is a conventional ferromagnet in the spin degrees of freedom. 

In the case when $\lambda$ is large, the spin parts of the Coulomb and ferromagnetic phases do not change, since both of these phases suppress hedgehogs. These phases are now trivial insulators in the boson degrees of freedom. On the other hand, in the paramagnetic phase the hedgehogs are proliferated and the bosons become bound to them. This is the binding phase, which we will argue is a topological phase protected by the $U(1)_{\rm spin}\times U(1)_{\rm boson}$ and $\ztwot$ symmetries.

\subsubsection{Symmetries When The Spins Are Represented By An Easy-plane \cp Model}
When representing our spins with an easy-plane \cp model we have explicitly broken the $SO(3)$ symmetry from the previous section down to a $U(1)$ symmetry which corresponds to spin rotations in the easy plane. 
This complicates our discussion of the discrete symmetries of the model. In the previous section all reflections of $\vec{n}$ were related to each other by an operation in $SO(3)$, but in the easy-plane case, e.g., reflections in the $ab$ plane $(n_a, n_b, n_c) \to (n_a, n_b, -n_c)$ are now distinct from reflections in the $ac$ plane $(n_a, n_b, n_c) \to (n_a, -n_b, n_c)$. The result of this is that there are now two $\ztwot$ symmetries which protect the topological phase, in the sense that as long as any one of these symmetries is preserved, the topological phase cannot be continuously connected to a trivial phase. 
The first $\ztwot$ symmetry is the same as that in the previous section; in the variables of Eq.~(\ref{sspin}) it reads:
\begin{equation}
\begin{array}{ccc}
 \phi_\uparrow&\rightarrow& -\phi_\downarrow \\
\phi_\downarrow&\rightarrow &-\phi_\uparrow \\
a_\mu&\rightarrow & -a_\mu \\
B_{\mu\nu}&\rightarrow & -B_{\mu\nu}\\
J_\mu &\rightarrow &-J_\mu \\
i & \rightarrow & -i
\end{array}.
\label{z2}
\end{equation}
This symmetry corresponds to a reflection in the $ab$ plane.

The second $\ztwot$ symmetry is a combination of reflections in the $ab$ and $ac$ planes and acts on spins as $(n_a, n_b, n_c) \to (n_a, -n_b, -n_c)$. In the \cp representation, this symmetry is given by:
\begin{equation}
\begin{array}{ccc}
 \phi_\uparrow&\rightarrow& \phi_\downarrow \\
\phi_\downarrow&\rightarrow &\phi_\uparrow \\
a_\mu&\rightarrow & a_\mu \\
B_{\mu\nu}&\rightarrow & B_{\mu\nu}\\
J_\mu &\rightarrow &J_\mu \\
i & \rightarrow & -i
\end{array}.
\label{z22}
\end{equation}
Note that a symmetry which consists only of a reflection in the $ac$ plane does not protect the topological phase, as under this symmetry we can still apply the Zeeman field in the $c$ direction and connect to a trivial phase as in Sec.~\ref{subsubsec:HeisSym}.  Note also that we cannot apply fields in the $ab$ plane as this would violate the $U(1)_{\rm spin}$ symmetry.

If either of the above $\ztwot$ symmetries are preserved, the topological phase cannot be connected to a trivial phase. A system with only the first symmetry has a total symmetry of $U(1)_{\rm spin}\times U(1)_{\rm boson}\times \ztwot$, while if we consider the second symmetry the direct product in front of $\ztwot$ is replaced with a semidirect product. (see Sec.~\ref{sec::discussion} for further discussion of these symmetries) The Zeeman field introduced in the previous section (and used again below) breaks both of these symmetries.

Note that in the presence of separate $U(1)_{\rm spin}$ and $U(1)_{\rm boson}$ symmetries, we could in principle consider unitary symmetries given by Eqs.~(\ref{z2}), (\ref{z22}), by simply omitting the complex conjugation.  However, if we imagine introducing a tunnelling between the two $U(1)$ symmetries, which would reduce the total symmetry to $U(1)\times\ztwot$ [or $U(1)\rtimes\ztwot$], the discrete symmetry should then treat both spin and boson variables in the same way. We find that this condition is satisfied as long as the above symmetries are understood as anti-unitary; hence we will refer to these symmetries as $\ztwot$.

\subsection{Observation of a Witten effect}
\label{subsec::witten}
The \cp representation allows us to make a bulk measurement which can detect whether our system is a bosonic topological insulator. This measurement, predicted in both the fermionic\cite{FranzWitten} and bosonic\cite{MaxWitten, Max, YeWen2014, YeWang2014} TIs, is called a Witten effect, and is the tendency of an external magnetic monopole in a TI to bind a charge of one-half. 
In order to justify our claim that the binding phase is a bosonic TI, we now demonstrate that our system exhibits a Witten effect. 

The first step in measuring a Witten effect in our Monte Carlo is to add external $U(1)$ gauge fields to the system. These external fields correspond to the $U(1)_{\rm spin}$ and $U(1)_{\rm boson}$ symmetries of the system. We will fix configurations for these fields before performing the simulations, which corresponds to putting our system in some external electromagnetic field. These external gauge fields are distinct from the internal, dynamical gauge field $a_\mu$. Similarly, the external monopoles introduced in this section are different from the hedgehogs (which are monopoles of the field $a_\mu$) discussed previously. We will introduce magnetic monopoles in the $U(1)_{\rm spin}$ gauge field and will measure $U(1)_{\rm boson}$ charge. The Witten effect is the statement that the external monopoles of the $U(1)_{\rm spin}$ gauge field will bind half of a charge of the $U(1)_{\rm boson}$ symmetry.

Let us first consider the gauge field coupled to the spin degrees of freedom. Here it is convenient to think in terms of a parton description. This is a description of a spin model by using the easy-plane \cp model in which the phases $\phi_\uparrow$ and $\phi_\downarrow$ represent the phases of different types of bosonic ``partons''. These partons each represent one-half of a physical boson. Each parton carries a unit charge under the internal gauge field $a$. The physical boson carries unit charge under the external gauge field, which we denote $A_1$. The partons carry half a charge under this gauge field, with one parton carrying positive charge and the other negative.
To modify Eq.~(\ref{sspin}) to reflect this, we add $\pm A_{1\mu}/2$ inside the cosines on the first line. 
Partially integrating out the parton fields then gives compact Maxwell terms in the field combinations $a_\mu+A_{1\mu}/2$ and $a_\mu-A_{1\mu}/2$, with equal couplings due to the $\ztwot$ symmetry. We can write each in the Villain form, which introduces two integer-valued placket variables $B^+_{\mu\nu}$ and $B^-_{\mu\nu}$. We can expand and recombine these quadratic terms to get separate terms for the $a$ and $A_1$ fields, leading to the following action:
\begin{eqnarray}
S&=&-\beta\sum_{R,\mu} \cos[\nabla_\mu\phi_\uparrow(R)-a_\mu(R)-\frac{1}{2}A_{1\mu}(R)]\nonumber\\
&-&\beta\sum_{R,\mu} \cos[\nabla_\mu\phi_\downarrow(R)-a_\mu(R)+\frac{1}{2}A_{1\mu}(R)]\nonumber\\
&+&\frac{K}{2}\sum_{R,\mu<\nu}\left[\nabla_\mu a_\nu(R)-\nabla_\nu a_\mu(R)-2\pi B_{\mu\nu}(R)\right]^2\nonumber\\
&+&\frac{K}{8}\sum_{R,\mu<\nu}\left[\nabla_\mu A_{1\nu}(R)-\nabla_\nu A_{1\mu}(R)-2\pi M_{\mu\nu}(R)\right]^2\nonumber\\
&+&\frac{\lambda}{2}\sum_{r,\mu} [ J_\mu(r)- Q_\mu(r)]^2+i\sum_{r,\mu}J_{\mu}(r)A_{2\mu}(r).
\label{withA}
\end{eqnarray}
Here $B_{\mu\nu}=(B^+_{\mu\nu}+B^-_{\mu\nu})/2$, and $M_{\mu\nu}=B^+_{\mu\nu}-B^-_{\mu\nu}$. Note that $B_{\mu\nu}$ and $M_{\mu\nu}$ are not completely independent variables but satisfy the conditions that $B_{\mu\nu}$ is $1/2$ $\times$ integer, $M_{\mu\nu}$ is integer, and 
\begin{equation}
2B_{\mu\nu}(R)=M_{\mu\nu}(R) ~~{\rm mod}~~2. 
\label{constraint}
\end{equation}
The variables $M_{\mu\nu}$ can be interpreted as the Dirac strings of the external monopoles. We can see that when we introduce external monopoles of odd integer strength, the internal hedgehog variables become half-integer valued---this is a crucial observation for the discussion of the Witten effect.\cite{Max}  Note that in the case of no external field $A_1$ and no external monopoles, the $B_{\mu\nu}$ variables are integer-valued and Eq.~(\ref{withA}) reduces to Eq.~(\ref{sspin}).  Note also that the coupling we wrote for $(\nabla_\mu A_{1\nu}-\nabla_\nu A_{1\mu}-2\pi M_{\mu\nu})^2$ [the ``Maxwell term'' for the external gauge field on the fourth line of Eq.~(\ref{withA})] is special to the preceding spinon-generated argument, while it is expected to be renormalized up by the rest of the universe; in fact, we will assume that the $A_1$ field is essentially externally controlled and is static.

We introduce a monopole into our system by making a specific choice for the external variables $A_{1\mu}$ and $M_{\mu\nu}$. First, we choose a configuration of $M_{\mu\nu}$ which will lead to a pair of external monopoles. In our system with periodic boundary conditions, it is not possible to have only a single monopole. We will place external monopoles at coordinates $(x,y,z)=(0,0,0)$ and $(0,0,L/2)$, on the lattice labelled by $r$. The external monopoles will have opposite charges, with the positively-charged one at the origin. All configurations of external gauge fields will be constant in the $\tau$ direction. In order to place external monopoles at these locations, we set $M_{xy}(R)=1$ whenever $X=-1/2$, $Y=-1/2$, and $1/2 \leq Z \leq L/2-1/2$ (the $1/2$'s come from the $R$ lattice being displaced from the $r$ lattice by half a lattice spacing). All other $M_{\mu\nu}$ are set to zero. By Eq.~(\ref{constraint}), we must also constrain $B_{xy}$ to be odd half-integers on this Dirac string. Now that we have specified the $M_{\mu\nu}$ values which introduce external monopoles, we choose values for the $A_{1\mu}$ variables to minimize the action of the Maxwell term on the fourth line of Eq.~(\ref{withA}). 

In our simulations we will set $A_{2\mu}=0$ everywhere, so that it does not affect the system. It will be used only when computing linear responses. 

There are in fact multiple configurations of the variables $M_{\mu\nu}$ which give the same configuration of external monopoles. The physics of the system is independent of which configuration of $M_{\mu\nu}$ we choose because the various configurations are related by the following gauge transformation:
\begin{equation}
\begin{array}{ccc}
M_{\mu\nu}(R)&\rightarrow&M_{\mu\nu}(R)+\nabla_\mu \kappa_\nu(R)-\nabla_\nu \kappa_\mu(R) \\
B_{\mu\nu}(R)&\rightarrow&B_{\mu\nu}(R)+\frac{1}{2}[\nabla_\mu \kappa_\nu(R)-\nabla_\nu \kappa_\mu(R)] \\
A_{1\mu}(R)&\rightarrow&A_{1\mu}(R)+2\pi \kappa_\mu(R)\\
a_\mu(R)&\rightarrow&a_\mu(R)+\pi \kappa_\mu(R)
\end{array},
\end{equation}
where $\kappa_\mu(R)$ is an integer-valued field living on the links of the lattice labelled by $R$. One can use Eqs.~(\ref{mondef}) and (\ref{withA}) to check that this transformation does not change the action, 
including the configurations and energetics of the external monopoles and gauge fields, so our results are independent of the specific choice of the Dirac string $M_{\mu\nu}$. 

Having introduced external monopoles into our system, we can begin to see why they should bind half a charge of the bosons. 
The argument goes as follows. First, when modifying the Dirac string variables $M_{\mu\nu}$ to insert external monopoles, we were also forced to modify the variables $B_{\mu\nu}$ in such a way as to introduce one-half of a hedgehog at the same locations as the external monopoles. However, we saw in the previous section that hedgehog-boson 'molecules', which have similar bare long-range interactions as just hedgehogs (i.e., carry hedgehog number), are proliferated in the binding phase. Therefore the $1/2$-hedgehog which we introduced will be screened by a ``cloud'' of hedgehog-boson molecules drawn from the rest of the system. This screening is analogous to Debye screening in a plasma. The screening cloud will carry a hedgehog number of one-half, but with opposite sign to the first hedghog, leading to a total hedgehog number of zero. Since in the binding phase hedgehogs are bound to charges, the cloud also carries a boson charge of one-half. Therefore we find that half of a boson charge has bound to the external monopole. We have tested this intuition by direct Monte Carlo simulations.

The above discussion is complicated by two degeneracies in Eq.~(\ref{withA}). First, there is a degeneracy between $Q_\tau(0,0,0,\tau)=1/2$ and $Q_\tau(0,0,0,\tau)=-1/2$: e.g.,~when $M_{xy}=1$ on the Dirac string, $B_{xy}$ can be either $+1/2$ or $-1/2$ with the same energy. This degeneracy is a result of the symmetry in Eq.~(\ref{z22}). In what follows it can be helpful to neglect variations in the $\tau$ direction and think about $Q_\tau(0,0,0,\tau)$ as a stationary hedgehog charge at the origin. Because of this degeneracy the statistical mechanics has each of the two states equally probable, which in an infinitely long simulation would lead to zero net hedgehog charge, and no observation of the Witten effect.
Second, if we were able to fix the hedgehog number in one of these two states, for example $Q_\tau(0,0,0,\tau) = 1/2$, we have another degeneracy, between $J_\tau(0,0,0,\tau) = 0$ and $J_\tau(0,0,0,\tau) = 1$. This leads to an average boson charge of $1/2$ at the location of the  monopole. This charge cancels the boson charge from the screening cloud, leading to no observation of the Witten effect. 

Despite these degeneracies, we may still observe a Witten effect if the degenerate states are metastable, and the Monte Carlo is stuck in one of the two states. For example, in order to get from $Q_\tau=+1/2$ to $Q_\tau=-1/2$ one needs to modify all of the $B_{\mu\nu}$ on the Dirac string, and the $\phi_{\uparrow,\downarrow}$ and $a_{\mu}$ variables nearby. Such a move would be quite unlikely (and impossible in the limit of an infinite separation between the monopoles). Similarly, to get from $J_\tau=0$ to $J_\tau=1$ one needs to insert a boson loop which passes from one monopole to another, and such a step is highly unlikely with local updates.

The results of our numerics shows that even at the small system sizes that we can access, degeneracy in the $J$ variables is always broken. However, the degeneracy in the $Q$ variables is unbroken, and so we do not observe a Witten effect. Let us consider the $Q$ degeneracy more carefully. One of the two degenerate states has $Q=+1/2$ bound to a positive external monopole at $r=0$, and $Q=-1/2$ bound to a negative external monopole at $r=L/2$. The other state as $Q=-1/2$ bound to a positive external monopole at $r=0$, and $Q=+1/2$ bound to a negative external monopole at $r=L/2$. Note that to change between the two degenerate states we need to modify variables along a string connecting the two external monopoles, and the probability of this happening reduces exponentially with distance, so in the thermodynamic limit this degeneracy would certainly break, even though on our small systems it does not. 

We will break this degeneracy in our system by decreasing the probability that the system will flip between degenerate states. To do this we add the following ``biasing'' term:
\begin{equation}
\delta S_{\rm bias}=\gamma  \sum_{\tau}\left[J_\tau(0,0,0,\tau)-J_\tau(0,0,L/2,\tau)\right],
\label{bias}
\end{equation}
where $\gamma$ is some small real number. We have scanned the system by increasing $\gamma$ from zero and seen no phase transitions, implying that these small $\gamma$ do not change the phase we are in. It would be surprising if the above biasing term affected the bulk physical properties of the system, as we are only making a modification to a fraction of the system proportional to $L^{-3}$. In addition, though this term breaks the symmetry in Eq.~(\ref{z2}), it preserves the symmetry in Eq.~(\ref{z22}), and the topological phase is protected as long as either of these symmetries is preserved. 

We hope that the above term reduces the probability of switching between degenerate states, so that we can observe a Witten effect. From our numerical results we can see that the above term indeed does this for a large range of $\gamma$. Therefore we have broken the problematic degeneracies and removed the obstacle to measuring the Witten effect. We would like to stress that the Witten effect is ordinarily defined for a single monopole, in the thermodynamic limit. The problems we have with degeneracies are artifacts of the fact that we are trying to measure a Witten effect in a finite-size system with two external monopoles. If the action were defined on an infinite system with only one monopole, these problems would not arise as it would take an infinitely long time for the system to change between degenerate configurations.

\begin{figure}
\includegraphics[angle=-90,width=\linewidth]{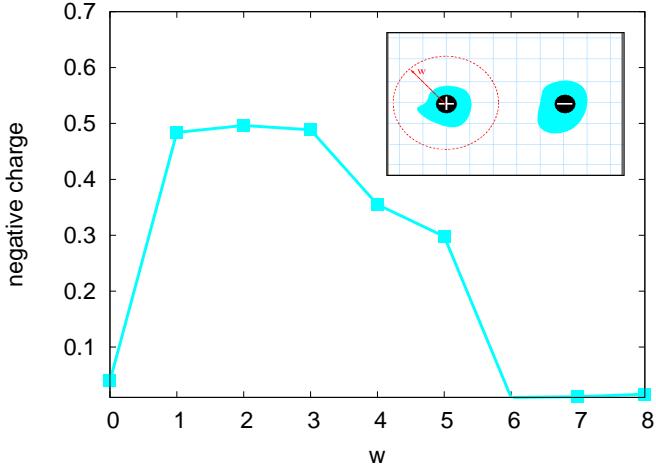}
\caption{Measurement of the Witten effect. The inset shows the measurement setup. External monopoles are inserted into the system, at a distance $L/2$ apart. They carry hedgehog numbers of $\pm 1/2$, and are Debye screened by hedgehogs of equal and opposite number, which also carry half of a boson charge. The main plot shows the boson charge enclosed in a sphere of radius $\scripty{r}$. For $\scripty{r}\approx 1-3$, this sphere measures the boson charge in the screening cloud near the origin, and the result is $1/2$, as expected. For $\scripty{r}\gtrsim 5$ the sphere includes the charge from the other screening cloud, and the enclosed charge drops to zero. The system size is $L=10$, using bulk parameters $\beta=0.2$, $K=0.4$, $\lambda=8$ (cf.\ bottom panel in Fig.~\ref{cp1bulkphase}), and local biasing parameter $\gamma=1.5$.}
\label{witten}
\end{figure}

We observe the Witten effect by measuring the total charge enclosed in a sphere of radius $\scripty{r}$, centered around the location of the monopole.  The precise definition of our measurement is:
\begin{equation}
{\rm charge}=\frac{1}{L}\sum_\tau \sum_{x^2+y^2+z^2\leq w^2} \langle J_\tau(x,y,z,\tau) \rangle ~.
\end{equation}
Note that the sphere discussed above is only a sphere in the $x$, $y$ and $z$ directions, and we have averaged over the $\tau$ direction.
We performed simulations with $L=10$ and show the results in Fig.~\ref{witten}.
At $\scripty{r}=0$, we are at the location of the monopole. There is nearly no boson charge bound here. At $\scripty{r}=1$, we have already included most of the screening cloud, and therefore measure an enclosed charge close to $1/2$, as expected. The fact that $\scripty{r}=1$ measures a value close to one half shows that the screening length in the system is quite short. At $\scripty{r}=2$, we have included the entire screening cloud, so the charge is even closer to $1/2$. As $\scripty{r}$ is further increased, we start to include the screening cloud from the other monopole, which is located at a distance $L/2$ from the first one. This cancels some of the charge from the first monopole, and so the total charge starts to decrease. When $\scripty{r}>L/2$, the sphere encompasses the screening clouds from both external monopoles, and so there is a total charge of nearly zero. The values of charge are negative because we set the biasing parameter $\gamma$ such that at the origin there is a monopole of positive charge, $Q=+1/2$, and the sign of the charge in the screening cloud is the opposite of the sign of the monopole. 

\begin{figure}
\includegraphics[angle=-90,width=\linewidth]{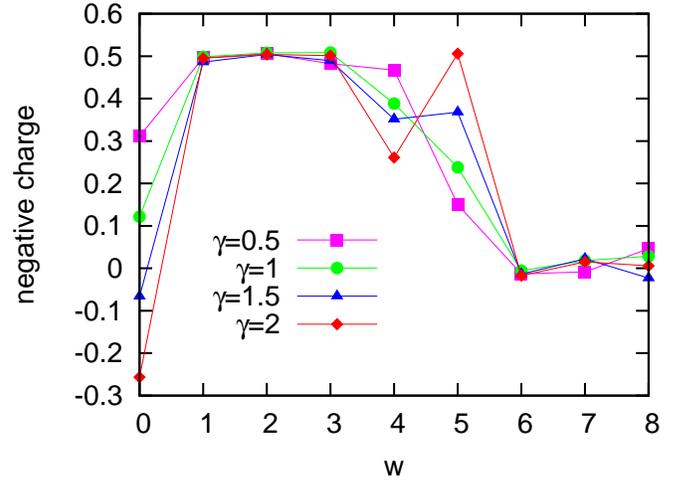}
\caption{Witten effect for different values of $\gamma$, but all other parameters the same as in Fig.~\ref{witten}. We see that near $\scripty{r}=2$, when we are measuring the charge inside a sphere which surrounds exactly one monopole, the amount of charge is approximately one-half, and independent of $\gamma$. At smaller $\scripty{r}$ (or at $\scripty{r} \gtrsim 4$ when the sphere starts overlapping with the screening cloud near the second monopole), the enclosed charge does depend on $\gamma$.}
\label{diffgamma}
\end{figure}

In Fig.~\ref{witten}, we have found that the amount of charge at the site of the monopole ($\scripty{r}=0$) is nearly zero. However, this is not universal and in fact depends on the choice of $\gamma$ in Eq.~(\ref{bias}). 
In our simulations, we find that $\scripty{r}=2$ is sufficiently far from the monopole to be unaffected by the change in $\gamma$. Figure~\ref{diffgamma} shows simulations taken with different values of $\gamma$. We see that though the amount of charge close to the monopoles (near $\scripty{r}=0$ and $\scripty{r}=L/2$) can be affected by changing $\gamma$, the value at $\scripty{r}=2$ is always very close to one-half, regardless of what $\gamma$ is used.

\begin{figure}
\includegraphics[angle=-90,width=\linewidth]{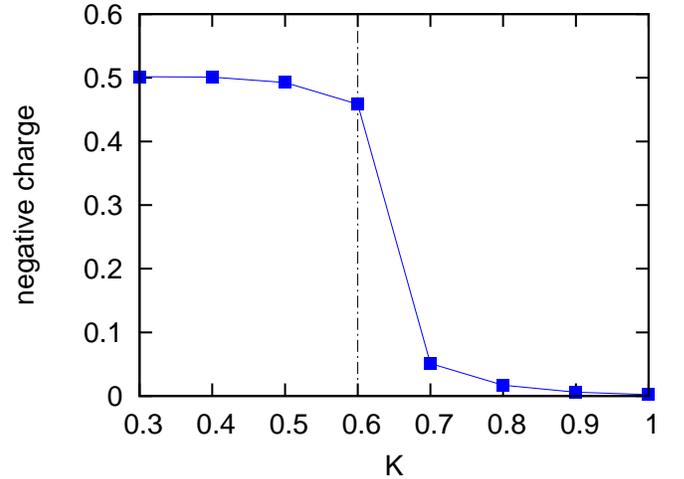}
\caption{A demonstration of how the Witten effect can be used to detect phase transitions. The plot shows the amount of charge enclosed by a sphere with $\scripty{r}=1$, while changing $K$ but keeping all other parameters the same as in Fig.~\ref{witten}. We can compare to Fig.~\ref{cp1bulkphase}, and see that at $K=0.6$, when the system transitions from the topological phase to the Coulomb phase, the 1/2-quantization of the enclosed charge abruptly stops.}
\label{wittenphase}
\end{figure}

Various other measurements can be made to support our conclusions. Measuring the total charge on each site shows that the half-charge is distributed around the monopole in an approximately spherically symmetric way. We can also use the Witten effect to detect phase transitions out of the bosonic TI. Figure~\ref{wittenphase} shows the total charge on all the nearest-neighbours, as a function of $K$. We note that the quantized Witten effect disappears at $K=0.6$, which is where the phase transition to the Coulomb phase is located, see Fig.~\ref{cp1bulkphase}. (In the Coulomb phase the amount of charge isn't necessarily zero, but it is not quantized and in our simulations we found it to be zero.) We also observe the disappearance of the bound charge when the system undergoes a transition to trivial insulator as $\eta$ is decreased to zero.

\subsection{Surface Phase Diagram}
\label{subsec::cp1surface}
The measurement of the Witten effect is evidence that our binding phase is a bosonic topological insulator. We can now study the exotic physics on the surface of this topological phase. We expect to find the surface phases predicted in Ref.~\onlinecite{SenthilVishwanath}.

We begin with some analytical arguments that provide a microscopic derivation of the surface field theory proposed in Ref.~\onlinecite{SenthilVishwanath}.
We define the surface as in Sec.~\ref{subsec:heissurf}. To uncover the exotic physics, we begin by performing a change of variables from the physical boson currents $J_\mu(r)$ to new integer-valued variables
\begin{equation}
G_\mu(r) \equiv J_\mu(r) - \eta(r) Q_\mu(r) ~,
\label{Gchange}
\end{equation}
which satisfy 
\begin{eqnarray*}
\left(\sum_\mu \nabla_\mu G_\mu\right) (x, y, z, \tau) &=& 
\delta_{z, z_R} Q_z(x, y, z_R-1, \tau) \\
&-& \delta_{z, z_L} Q_z(x, y, z_L-1, \tau) ~. 
\end{eqnarray*}
The action for the spins and the new $G_\mu(r)$ variables is simply
\begin{equation}
S = S_{\rm spin} + \sum_{r,\mu} \frac{\lambda_\mu(r)}{2} \Big[G_\mu(r)\Big]^2 ~.
\end{equation}

For simplicity, from now on we consider situation where the trivial phase region and the SPT phase region are deep in their respective phases: in particular, $\lambda_{\rm bulk}$, defined as in Eq.~(\ref{bulkvsurf}), is very large.  At first we further simplify the situation by taking $\lambda_{\rm surf}$ to be very large, in which case we expect the variables $G_\mu(R)$ to be zero everywhere.  We also assume that $\beta$ is small everywhere, so the spin variables want to be deep in the disordered phase.  However, near the two surfaces, the spin configurations must satisfy
\begin{equation}
\begin{array}{c}
Q_z(x, y, z_R-1, \tau) = 0, \\
Q_z(x, y, z_L-1, \tau) = 0.
\end{array}
\end{equation}
Focusing on the spins near one surface, say at $z_R$, we can view $Q_z(x, y, z_R-1, \tau)$ as simply hedgehog numbers in the corresponding (2+1)D spin system spanned by sites $(X, Y, Z=z_R-1/2, T)$, and the above conditions correspond to complete suppression of hedgehogs in this spin system.  Such a (2+1)D Heisenberg $O(3)$ spin model with hedgehog suppression was studied in Ref.~\onlinecite{LesikAshvin} and argued to be described by a \emph{non-compact} $CP^1$ field theory ($NCCP^1$).  On a simple (2+1)D cubic lattice, the Heisenberg model with complete hedgehog suppression actually has spontaneous magnetic order of spins even when the direct spin-spin interactions are zero.\cite{LauDasgupta, KamalMurthy}  However, more generic such models can have a spin-disordered phase with a propagating ``photon,''\cite{KamalMurthy, LesikAshvin} as well as other phases such as coexistence of the magnetic order and the propagating photon.\cite{LesikAshvin2}  We will see that our findings in the present simulations on the surface of the bosonic TI region are consistent with these earlier results.

Let us now proceed more systematically and, in particular, show how we obtain a generic $NCCP^1$ model on the surface of the bosonic TI region.  For simplicity, we take $\lambda_{\rm bulk}$ to be very large. For finite $\lambda_{\rm surf}$, we need to keep $G_x, G_y, G_\tau$ degrees of freedom in the (2+1)D ``layer'' at $z_R$, while all other $G_\mu(r)$ are zero.  Focusing on the spin variables residing on sites $(X, Y, Z=z_R-1/2, T)$, the hedgehogs in this (2+1)D system are given precisely by $Q_z(x, y, z_R-1, \tau)$, which we will denote simply as $Q(x, y, \tau)$.  The structure of the surface theory is
\begin{eqnarray}
S_{\rm surface} &=& S_{\rm matter-gauge} + \frac{K}{2}\sum  ({\bm\nabla} \times {\bm a} - 2\pi {\bm B})^2\nonumber \\
&+& \frac{\lambda_{\rm surf}}{2}\sum  {\bm G}^2 ~,
\label{Ssurf}
\end{eqnarray}
subject to constraints
\begin{equation}
 \nabla_x G_x + \nabla_y G_y + \nabla_\tau G_\tau \equiv {\bm \nabla} \cdot {\bm G} = Q(x,y,\tau) \equiv {\bm \nabla} \cdot {\bm B} ~.
\end{equation}
Here $S_{\rm matter-gauge}$ represents the first term in Eq.~(\ref{sspin}) restricted to the surface degrees of freedom. The above is a 3D statistical mechanics model, and $a_\mu$, $(\nabla\times a)_{\mu\nu} \equiv \nabla_\mu a_\nu -\nabla_\nu a_\mu$, and $B_{\mu\nu}$ from Eq.~(\ref{sspin}) can be now defined as 3-vectors and are denoted by bold-face (e.g., $\mu$-th component of ${\bm B}$ is $\frac{1}{2}\epsilon_{\mu\nu\rho}B_{\nu\rho}$). We have suppressed position indices to simplify notation.

This surface theory has spins plus integer-valued ``currents'' $G_\mu$ sourced and sinked by the hedgehogs of the spin system.  When the ``line tension'' $\lambda_{\rm surf}$ for the lines formed by these ``currents'' is large, we intuitively expect that the hedgehogs of the spin system are linearly confined.  It is not immediately clear, however, what happens when $\lambda_{\rm surf}$ is small.  Below we argue that the surface is still qualitatively described by the same ``hedgehog-suppressed'' field theory, which, however, can be in different regimes and have several different phases.

 We can deal with the constraints in the partition sum by changing to new variables
\begin{equation}
{\bm B}^\prime = {\bm B} - {\bm G} ~,
\end{equation}
which satisfy
\begin{equation}
{\bm \nabla} \cdot {\bm B}^\prime = 0 ~.
\end{equation}
The action becomes 
\begin{eqnarray*}
S_{\rm surface} &=& S_{\rm matter-gauge} + \frac{K}{2}\sum  ({\bm\nabla} \times {\bm a}  - 2\pi {\bm B}^\prime - 2\pi {\bm G})^2 \\
 &+& \frac{\lambda_{\rm surf}}{2}\sum  {\bm G}^2 ~.
\end{eqnarray*}
There are now no constraints on the ${\bm G}$ variables, and we can formally sum these out to obtain a local action which is a function of ${\bm\nabla} \times {\bm a} - 2\pi {\bm B}^\prime$,
\begin{eqnarray}
S_{\rm surface} &=& S_{\rm matter-gauge} + S_{\rm gauge, eff}[{\bm\nabla} \times {\bm a}  - 2\pi {\bm B}^\prime] .
\label{partsurface}
\end{eqnarray}
However, any such action with the compact variables ${\bm a}$ and divergenceless, integer-valued ${\bm B}^\prime$ can be formally viewed as describing a non-compact gauge field!  In the limit of large $\lambda_{\rm surf}$, the effective action will have essentially lattice Maxwell form with stiffness $K$, while for intermediate to small $\lambda_{\rm surf}$ the gauge field energy will have more complicated but still local form.  Thus, the field theory at the surface has the spinon matter fields coupled to a non-compact gauge field.  In particular, we expect that the surface can be in the same phases as the (2+1)D easy-plane $NCCP^1$ model.

We can also study how the surface action is coupled to the external gauge fields introduced in Eq.~(\ref{withA}). The minimal coupling between $J_\mu$ and $\Aext_{2\mu}$, combined with the change of variables in Eq.~(\ref{Gchange}), leads to the following term:
\begin{equation}
i \sum_{r,\mu} G_\mu(r) \Aext_{2\mu}(r) + i \sum_{r,\mu} \eta(r) Q_\mu(r) \Aext_{2\mu}(r) ~.
\end{equation}
It is convenient here to represent $Q_\mu(r)$ as
\begin{equation}
Q_\mu(r) = \frac{1}{2} \epsilon_{\mu\nu\rho\sigma} \nabla_\nu 
\left(B_{\rho\sigma} - \frac{\nabla_\rho a_\sigma - \nabla_\sigma a_\rho}{2\pi} \right) ~,
\end{equation}
where we have added a formal zero to the defining Eq.~(\ref{mondef}).
Using this in the preceding equation and integrating the second term by parts, we find both an additional bulk term as well as a surface term which results from taking a derivative of $\eta(r)$. 
Focusing again on the (2+1)D layer at $z_R$, we can write the surface contributions as
\begin{equation}
i \sum \left({\bm G} + \frac{{\bm\nabla} \times {\bm a}}{2\pi} - {\bm B} \right) \cdot {\bm A}_2 = 
i \sum \frac{{\bm\nabla} \times {\bm \alpha}}{2\pi} \cdot {\bm A}_2 ~,
\label{surfaceCS} 
\end{equation}
where we defined
\begin{eqnarray}
{\bm\nabla} \times {\bm \alpha} \equiv {\bm\nabla} \times {\bm a} - 2\pi {\bm B}^\prime ~,
\end{eqnarray}
which is precisely the flux of the {\it non-compact} gauge field identified in Eq.~(\ref{partsurface}).
When this is combined with Eq.~(\ref{partsurface}), we are left with an effective action for the surface with schematic Lagrangian density
\begin{equation*}
\left|\left( {\bm \nabla} - i {\bm \alpha} \mp i\frac{{\bm A}^{\rm ext}_1}{2} \right) z_{\up/\dn} \right|^2 + \frac{\kappa}{2} ({\bm \nabla} \times {\bm \alpha})^2 + i \frac{{\bm\nabla} \times {\bm \alpha}}{2\pi} \cdot {\bm A}_2 ~.
\end{equation*}
This action, which we derived from our lattice model, has precisely the easy-plane \nccp form proposed in Ref.~\onlinecite{SenthilVishwanath}. In claiming that this is the correct effective action of the surface, we have neglected bulk terms which in general may also contribute to the surface response properties. We do not have an analytical justification for this choice, though from the Monte Carlo study presented in Sec.~\ref{cp1Hall} we find that essentially only the surface terms given above contribute to the measured response properties, and it seems plausible that our argument applies in the limit of a sharp boundary between the topological and trivial phases deep in their respective regimes.

Note that the above arguments were based on the assumption that the $U(1)$ and $\ztwot$ symmetries were preserved in the bulk. If the $U(1)$ symmetry is broken in the bulk, the entire derivation of Eq.~(\ref{Ssurf}) based on conserved currents is invalid.  On the other hand if only the time reversal is broken in the bulk, the derivation naively holds, but in the matter-gauge sector there is no reason for $z_{\up}$ and $z_{\dn}$ to enter symmetrically---in particular, no reason for them to carry precise $+1/2$ and $-1/2$ charges, and the field theory written above is not valid. (In fact, the system will have non-quantized $\sigma_{xy}$ proportional to the length of the system in the z-direction). Since when the symmetry is broken the bulk ceases to be a topological phase, we of course should not expect exotic physics on the surface in this case.

\begin{figure}
\includegraphics[angle=-90,width=\linewidth]{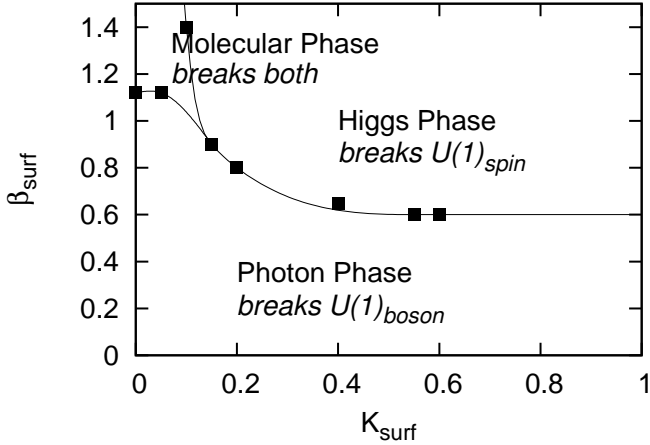}
\caption{Surface phase diagram on the boundary of the SPT phase in the model with a $CP^1$ version of the spins. The bulk parameters are $\beta_{\rm bulk}=0.2$, $K_{\rm bulk}=0.2$, $\lambda=8$, and the surface parameters are varied. The phase diagram has the same structure as the one found in the $NCCP^1$ model in Ref.~\onlinecite{LesikAshvin2}. The phases are the same as those in Fig.~\ref{heissurf}, though in this figure we have replaced the $SO(3)$ symmetry with $U(1)_{\rm spin}$, and the $\beta_{\rm surf}$ axis is oriented horizontally in Fig.~\ref{heissurf} and vertically in the present figure. }
\label{cp1surfphase}
\end{figure}

We can confirm the above arguments, which were made in some simplifying limits, by studying the system in Monte Carlo. We can determine the phase diagram of the surface by looking at singularities in the heat capacity. We can also study the magnetization and current-current correlators as described in the previous section. In this phase diagram we set the bulk parameters so that the system is in the topological phase; specifically, we take $\beta_{\rm bulk}=0.2$, $K_{\rm bulk}=0.2$, and $\lambda=8$ (cf.\ bottom panel of Fig.~\ref{cp1bulkphase}). We then vary the surface parameters and obtain the phase diagram shown in Fig.~\ref{cp1surfphase}, which is in good agreement with the phase diagram of the $NCCP^1$ model in the literature. Labels on the phase diagram are taken from Ref.~\onlinecite{LesikAshvin2}, and their relation to labels in Fig.~\ref{heissurf} is described below.
There are three phases in the diagram. At small $\beta_{\rm surf}$ the $\phi_{\uparrow, \downarrow}$ variables are disordered, conserving the $U(1)_{\rm spin}$ symmetry; the $U(1)_{\rm boson}$ symmetry is broken and the corresponding Goldstone mode is precisely the propagating photon in the \nccp theory on the surface, hence the label ``Photon Phase'' in Fig.~\ref{cp1surfphase}. 
At large $\beta_{\rm surf}$, $K_{\rm surf}$ the partons $z_\uparrow$, $z_\downarrow$ are condensed, leading to a ``Higgs Phase'' (in the \nccp language) in which the $U(1)_{\rm spin}$ symmetry is broken but the $U(1)_{\rm boson}$ symmetry is preserved. 
Finally at large $\beta_{\rm surf}$ and small $K_{\rm surf}$ both the $U(1)_{\rm spin}$ and $U(1)_{\rm boson}$ symmetries are broken.  In the \nccp language, the individual $z_\uparrow$ and $z_\downarrow$ are gapped so the gauge field $a_\mu$ is free to fluctuate, but the ``molecular field'' $\Psi_{\rm mol} \sim z_\uparrow^\dagger z_\downarrow$, which is precisely the easy-plane spin field, $\Psi_{\rm mol} \sim n_a + in_b$, becomes ordered, hence the label ``Molecular Phase'' in Fig.~\ref{cp1surfphase}.
We emphasize that the microscopic model we are simulating in (3+1)D has a compact gauge field, and we are detecting the presence or absence of $U(1)_{\rm spin}$ and $U(1)_{\rm boson}$ symmetry breaking on the surface by direct measurements.  It is remarkable that the surface phase diagram is captured by the \nccp field theory with \emph{non-compact} gauge field!

All of the phases in Fig.~\ref{cp1surfphase} break either a $U(1)_{\rm spin}$ or a $U(1)_{\rm boson}$ symmetry and are therefore superfluids. As in the previous section, without access to the properties of their gapped excitations we cannot directly confirm that they are the predicted surface phases. As in the previous section, our phase diagram contains a direct transition between the superfluid phases, which can be viewed as providing some evidence for the proposed surface physics and is also predicted to be a deconfined critical point.\cite{SenthilVishwanath}

\subsection{Symmetric Surface Phase with Topological Order}
\begin{figure}
\includegraphics[angle=-90,width=\linewidth]{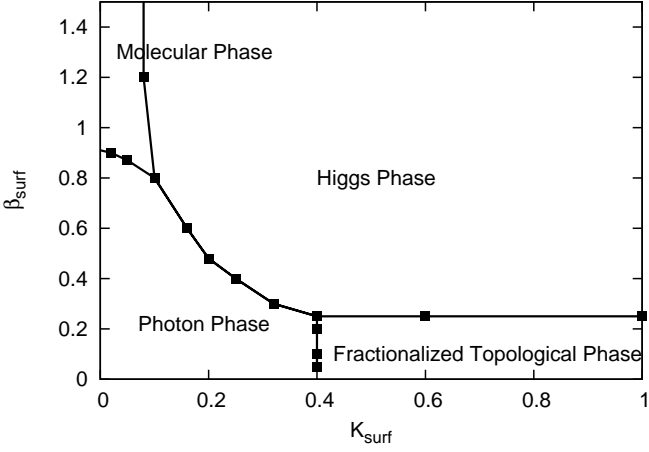}
\caption{Surface phase diagram for the easy-plane $CP^1$ version of the model, with the additional pairing term on the surface given by Eq.~(\ref{pairing}). This diagram was obtained for $\beta_{\rm bulk}=0.2$, $K_{\rm bulk}=0.2$, $\lambda=8$, and on the surface $t_{\rm pair}=2$, $\beta_{\rm surf}$ and $K_{\rm surf}$ varied. Compared to Fig.~\ref{cp1surfphase}, we see that there is a new phase at small $\beta_{\rm surf}$ and large $K_{\rm surf}$.  We expect that this surface phase is fully symmetric and has intrinsic topological order.
}
\label{cp1surfpair}
\end{figure}

Vishwanath and Senthil proposed that it is also possible to have a surface phase which is gapped and breaks no symmetries but has intrinsic topological order.\cite{SenthilVishwanath} Since this phase is not featured in Fig.~\ref{cp1surfphase}, we need to add another term to our surface action in order to push the system into this phase. The term we need to add is the following parton ``pair hopping term:''\cite{SenthilVishwanath, Max}
\begin{equation}
S_{\rm pair} = -t_{\rm pair}\sum_{R,\mu} \cos[\nabla_\mu (\phi_\uparrow + \phi_\downarrow) - 2 a_\mu],
\label{pairing}
\end{equation}
where we have included proper coupling to the gauge fields.  Note particularly that the pair field $\Psi_{\rm pair} \sim z_\uparrow z_\downarrow$ does not carry $U(1)_{\rm spin}$ charge. 
We can now see what happens to the surface phase diagram (Fig.~\ref{cp1surfphase}) when we increase $t_{\rm pair}$ trying to induce condensation of $\Psi_{\rm pair}$.  For sufficiently large $t_{\rm pair}=2$, we get the phase diagram in Fig.~\ref{cp1surfpair}.  We see that a new phase has opened up at small $\beta_{\rm surf}$ and large $K_{\rm surf}$, where, as we will argue, $\Psi_{\rm pair}$ is condensed while the individual $z_\uparrow$ and $z_\downarrow$ are gapped.

When all the couplings $\beta_{\rm surf}$, $K_{\rm surf}$, and $t_{\rm pair}$ are small, there is nothing which can order the spins or gap the bosons. Therefore we are in the photon phase, which conserves $U(1)_{\rm spin}$ and breaks $U(1)_{\rm boson}$. To get a pairing phase with no broken symmetries, we need to restore the $U(1)_{\rm boson}$ symmetry without breaking the $U(1)_{\rm spin}$ symmetry, which can be acheived by condensing $\Psi_{\rm pair}$.  Let us first recall how the various terms in the action change the system.  The $\beta_{\rm surf}$ term allows hopping of the partons $z_{\uparrow,\downarrow}$, but even when this hopping is strong the fluctations in the gauge field $a_\mu$ when $K_{\rm surf}$ is small prevent the partons from condensing. The combination $\Psi_{\rm mol}\sim z_\uparrow^\dagger z_\downarrow$ can condense, and this breaks $U(1)_{\rm spin}$ and takes us to the molecular phase.
We can see from Fig.~\ref{cp1surfphase} that the $K_{\rm surf}$ term on its own does not change the phase of the system if $\beta_{\rm surf}$ is kept small. However, when it is combined with the $\beta_{\rm surf}$ term it can prevent fluctations in the $a_\mu$ field. This gaps the photon, and allows $z_\uparrow$ and $z_\downarrow$ to condense. This gives us the Higgs phase, where the $U(1)_{\rm boson}$ symmetry has been restored, but the $U(1)_{\rm spin}$ symmetry has been broken.

With this in mind, we can see why the $t_{\rm pair}$ term brings us into the topological phase. The $t_{\rm pair}$ term is similar to the $\beta_{\rm surf}$ in that it is also a hopping term, though it hops pairs of partons. Therefore when both $t_{\rm pair}$ and $\beta_{\rm surf}$ are large the two terms cooperate, which is why the $U(1)_{\rm spin}$ symmetry breaks at lower $\beta_{\rm surf}$ in Fig.~\ref{cp1surfpair} than in Fig.~\ref{cp1surfphase}. However, when $\beta_{\rm surf}$ is absent, the $t_{\rm pair}$ only hops pairs of spinons, and so it can condense $\Psi_{\rm pair}$ without condensing the individual $z_{\up,\dn}$ or $\Psi_{\rm mol}$. When the $t_{\rm pair}$ term is combined with the $K_{\rm surf}$ term the fluctuations of $a_\mu$ are gapped. Therefore the phase at large $t_{\rm pair}$, large $K_{\rm surf}$, and small $\beta_{\rm surf}$ can restore $U(1)_{\rm boson}$ without breaking $U(1)_{\rm spin}$, and this is the phase we are looking for.

We have confirmed that the pairing phase breaks neither $U(1)$ symmetry by direct measurements in the spin and boson sectors.  Also,  $\Psi_{\rm pair}$ is invariant under the $\ztwot$ symmetry in Eq.~(\ref{z2}), so this symmetry is not broken either, and we indeed do not observe any Hall response on the surface.  Therefore we believe that this phase is the fully symmetric gapped phase envisioned by Vishwanath and Senthil, which they argued has intrinsic topological order.  Specifically, we expect to have gapped spinon excitations carrying 1/2 of the $U(1)_{\rm spin}$ charge; at the same time, we also have vortices in the $\Psi_{\rm pair}$ field which carry 1/2 of the unit flux of the $a_\mu$ gauge field and hence 1/2 of the $U(1)_{\rm boson}$ charge; finally, the spinons and the vortices in $\Psi_{\rm pair}$ clearly have mutual statistics of $\pi$.  Unfortunately, we do not have simple direct measurements to confirm the topological order on the surface.  However, the indirect evidence for this scenario is very strong.

Thus, it is suggestive to compare Fig.~\ref{cp1surfpair} with the phase diagram obtained in Fig.~1 of Ref.~\onlinecite{Loopy}. In that work, we studied a (2+1)D model with $U(1)\times U(1)$ symmetry and mutual statistics between two different species of bosons. When the mutual statistics is $\pi$, we get a phase diagram with the same topology as Fig.~\ref{cp1surfpair}. The phase diagram contains a topological phase, two phases where one of the $U(1)$ symmetries is broken, and a phase where both symmetries are broken. There is a direct transition between the phases with one broken symmetry, which, if it were continuous, is a candidate for a deconfined critical transition.\cite{Gen2Loops} The surface of our bosonic topological phase is thought to have a similar field theory to that in our previous work,\cite{Loopy,Gen2Loops} and so we expect that the interpretations of the phases and phase transitions are the same in both models. We also remark that in Ref.~\onlinecite{FQHE} we presented a microscopic local Hamiltonian which has a topological phase with the same content of excitations. However, that phase also breaks $\ztwot$ symmetry and in particular has $\sigma_{xy}=1/2$, while the present surface phase respects $\ztwot$ and has no Hall response.

\subsection{Time-Reversal Breaking and Hall Effect on the Surface}
\label{cp1Hall}

Vishwanath and Senthil\cite{SenthilVishwanath} predict another exotic phase on the surface of the bosonic topological insulator---a phase which breaks the $\ztwot$ symmetry and has a Hall conductivity quantized to an odd integer (in units of $\frac{e^2}{h}$).  We can test this prediction in our Monte Carlo simulations. The first step is to break the $\ztwot$ symmetry on the surface. By examining Eq.~(\ref{z2}), we see that one way to break the symmetry is to replace the parameter $\beta$ in Eq.~(\ref{sspin}) by the parameters $\beta_\uparrow$ and $\beta_\downarrow$, which appear in the terms containing $\phi_\uparrow$ and $\phi_\downarrow$ respectively. When $\beta_\uparrow \neq \beta_\downarrow$, the $\ztwot$ symmetry is broken; this is roughly like applying the Zeeman field in Sec.~\ref{subsubsec:HeisSym}.

We start with $K=0.4$ and $\lambda=8$ everywhere, $\beta_{\rm bulk}=0.2$, and $\beta_{\uparrow}=\beta_{\downarrow}=0.2$ on the surface. This system will have its bulk in the topological phase, and its surface in the photon phase. We break the $\ztwot$ symmetry on one of the surfaces by increasing $\beta_{\uparrow}$. We expect that a small increase will not change the properties of the system very much, since $z_\uparrow\sim e^{i\phi_\uparrow}$ and $z_\downarrow\sim e^{i\phi_\downarrow}$ will still be gapped.\cite{LesikAshvin} However, as $\beta_\uparrow$ is further increased, $z_\uparrow \sim e^{i\phi_\uparrow}$ ``condenses'' and vortices in the $\phi_\uparrow$ variables will become gapped. In our simulations we see a singularity in the specific heat measured on the surface, indicating that the system has entered a new phase. We expect that this is the phase that will have Hall conductivity quantized at odd integer. In our simulations we also break time-reversal symmetry in the opposite direction on the other surface by increasing $\beta_\downarrow$.  As discussed in Sec.~\ref{subsec:heissurf}, in this setup the top and bottom surface taken together have Hall conductivity adding to two.

 We can see that unlike in the Heisenberg model, Eq.~(\ref{sspin}) is a differentiable function of the probing fields $A_1$ and $A_2$, and so we know how to properly couple the external gauge fields and can use linear response theory to compute the Hall conductivity. If our system has a non-zero Hall conductivity, then we can imagine integrating out the internal degrees of freedom to get the following effective action in terms of the external fields at the surface:
\begin{equation}
S_{\rm eff} = i \sum_{\rm surface} 
\frac{\sigma_{xy}^{12}}{4\pi} [{\bm A}_1 \cdot ({\bm \nabla} \times {\bm A}_2) + {\bm A}_2 \cdot ({\bm \nabla} \times {\bm A}_1)] ~,
\label{Seff}
\end{equation}
where bold face denotes three-component vectors appropriate for the (2+1)D surface, e.g., ${\bm A}_1 = (A_{1x}, A_{1y}, A_{1\tau})$, and the above form specifies our convention for $\sigma^{12}_{xy}$ (these units are such that $e^2/h=1$).  By taking, e.g., ${\bm A}_1 = (A_{1x}, 0, 0)$ and ${\bm A}_2 = (0, A_{2y}, 0)$, we have
\begin{eqnarray}
S_{\rm eff} &=& -i \sum_{\rm surface} \frac{\sigma_{xy}^{12}}{2\pi} A_{1x} \nabla_\tau A_{2y} ~.
\end{eqnarray}
Going to momentum space, we can obtain the Hall conductivity by:
\begin{equation}
\sigma_{xy}^{12}(k) = \lim_{A_1,A_2 \to 0} \frac{2\pi}{2 \sin(\frac{k_\tau}{2})}\frac{\partial^2 \ln Z}{\partial A_{1x}(k) \partial A_{2y}(-k)} ~,
\end{equation}
where $Z$ is the partition sum, and we also took $k = (0, 0, k_\tau)$.  Note that ${\bm A}_1$ and ${\bm A}_2$ reside on lattices dual to each other, and when defining the Fourier transforms we take the convention to transform in the absolute coordinates of the origins of the links (namely, lattice sites on the dual lattice have absolute coordinates displaced from the direct lattice by half of lattice spacing). 
Starting from the microscopic model, we can evaluate this conductivity from the current-current correlation functions:
\begin{equation}
\sigma_{xy}^{12}(k)=\frac{2\pi}{2\sin(\frac{k_\tau}{2})} \left\langle\xi_x(-k) J_y(k) \right\rangle,
\end{equation}
where the $U(1)_{\rm spin}$ current on a link $R,\mu$ is
\begin{equation}
\xi_\mu(R)\equiv\frac{i}{2}\left[\beta_\uparrow\sin(\nabla_\mu\phi_\uparrow-a_\mu)-\beta_\downarrow\sin(\nabla_\mu\phi_\downarrow-a_\mu)\right]\nonumber ~.
\end{equation}
The measurements are performed at the smallest wave-vector $k_{\rm min}=(0,0, 2\pi/L)$, as described in Section~\ref{subsec::bulkheis}. 

Note that the above conductivity measures the response of the $U(1)_{\rm spin}$ currents to applied fields coupled to the bosons (or vice-versa). To study SPTs with single $U(1)$, we can take the usual approach in the literature\cite{SenthilVishwanath} and ``glue'' the $U(1)_{\rm spin}$ and $U(1)_{\rm charge}$ by identifying $A_1$ and $A_2$ in Eq.~(\ref{Seff}); the conventional definition of $\sigma_{xy}$ in the case with single $U(1)$ is then related to the above $\sigma_{xy}^{12}$ via $\sigma_{xy} = 2 \sigma_{xy}^{12}$.
In particular, when the gauge fields are identified the Hall conductivity of a two-dimensional system of bosons is quantized to $2$ times an integer (in units of $e^2/h$), while the Hall conductivity on the surface of a topological phase is an odd integer. Therefore where we present numerical values we show $2 \sigma^{12}_{xy}$ so that the results can be easily compared to the literature values.

\begin{figure}
\includegraphics[width=\linewidth]{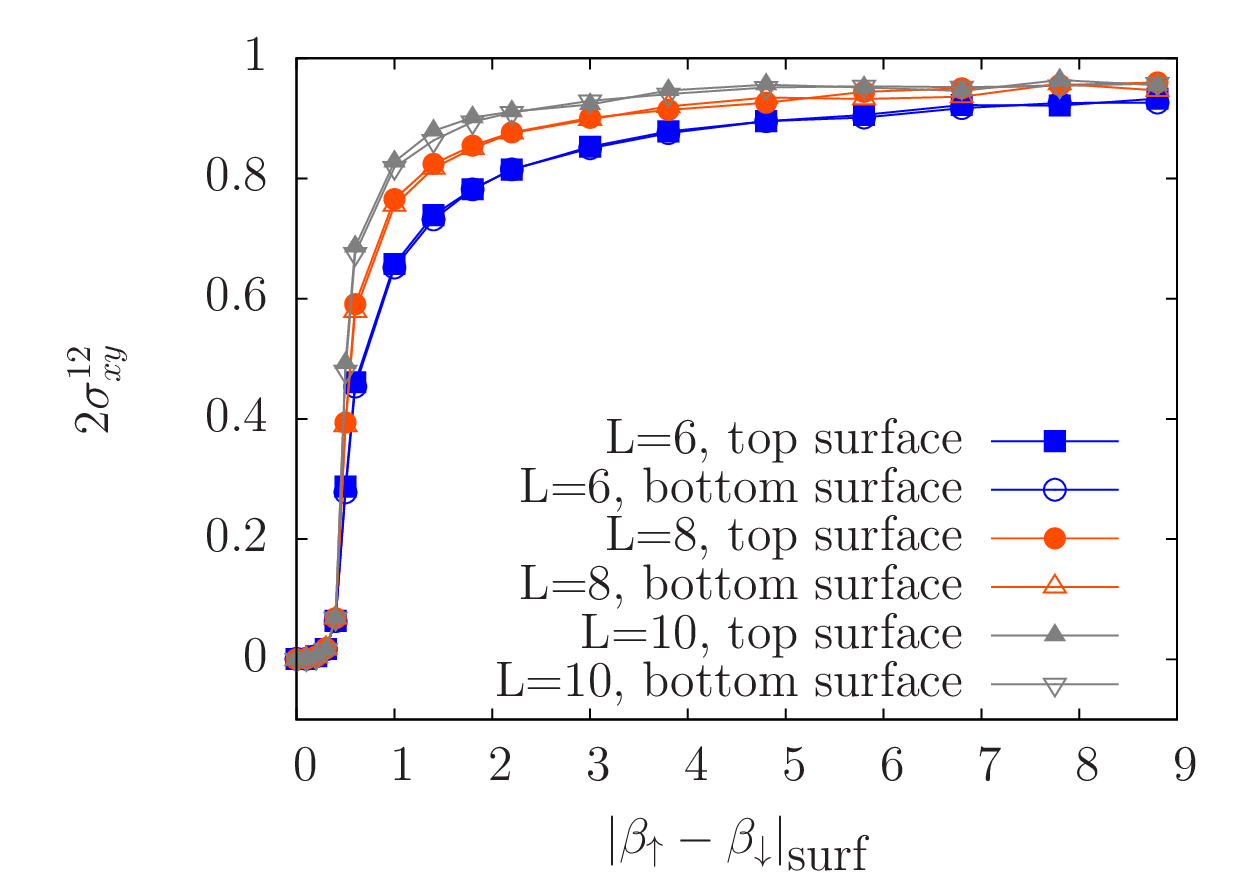}
\caption{Hall conductivity on the surface of the binding phase realized in the $CP^1$ version of the model, measured in units of $e^2/h$ (see text for details). On each surface we find that the conductivity is quantized to $1$, and this odd-integer value shows that we are on the surface of a bosonic TI. Data was taken for $K=0.4$, $\lambda=8$, $\beta_\downarrow=\beta_{\rm bulk}=0.2$. We measure the same conductivity on both the top and bottom surfaces of the topological phase.
}
\label{cp1hall}
\end{figure}

Figure \ref{cp1hall} shows our numerical measurements of the Hall conductivity. The horizontal axis is the strength of the $\ztwot$ symmetry breaking, which we will loosely call Zeeman field. We see that initially there is no quantized Hall conductivity, until the Zeeman field is strong enough to forbid one species of vortex (realized here by ``condensing'' the corresponding spinon species). At this point the Hall conductivity increases and reaches the value of approximately $1$. Though the value observed is actually slightly less than $1$, we believe that a large part of this is a finite-size effect, and indeed as the system size is increased the Hall conductivity gets closer to the expected value.  Note that we performed this measurement at precisely $z=z_R$.  It is {\em a priori} possible that the Hall conductivity could be spread among several values of $z$ near the surface, but this spreading is apparently very small, so that including additional layers does not change the result. In addition to the plot shown, we have performed the same measurement for several different values of $K$, $\beta$ and $\lambda$, and found that the quantized result is independent of these parameters as long as the bulk stays in the topological phase. This odd integer cannot be observed in a purely two-dimensional system with only short-ranged entanglement, and therefore this observation shows that we are measuring the Hall conductivity on the surface of a bosonic TI.

\section{Realizing symmetry-enriched topological phases by binding multiple hedgehogs to a boson}
\label{section::multiple}

In all of the above sections, we have studied a system where a single boson is bound to a single hedgehog. In this section we will describe the new physics which arises when our system contains bound states of a boson and multiple hedgehogs. We induce such a binding by making the following modification to Eq.~(\ref{action}):
\begin{equation}
S=S_{\rm spin}+\frac{\lambda}{2}\sum_{r,\mu} [ dJ_\mu(r)- Q_\mu(r)]^2.
\label{cdbind}
\end{equation}
Here $d$ is an integer, and for large $\lambda$ the action will bind a boson to $d$ hedgehogs, since the $\lambda$ term is minimized by $(Q, J) = (d, 1) \times~{\rm integer}$. 

\begin{figure}
\includegraphics[angle=-90,width=0.9\linewidth]{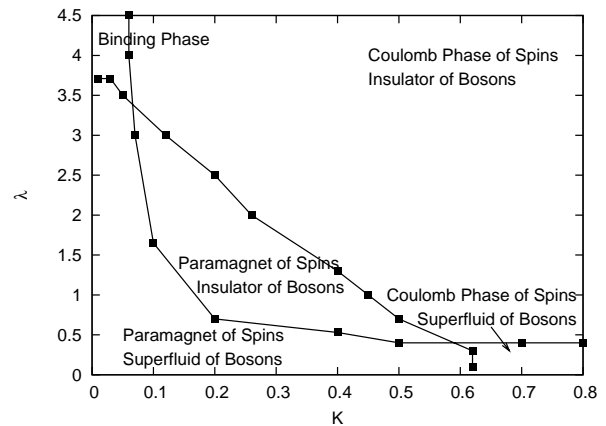}
\caption{Bulk phase diagram for the model described in Eqs.~(\ref{sspin}) and (\ref{cdbind}), with $d=3$ and small $\beta=0.1$. We can compare this to the case of $d=1$, where the middle phase is absent and the phase boundaries are straight vertical and horizontal lines; and to $d=2$, where the middle phase is absent and there is a line of phase transitions between the paramagnet/superfluid phase in the lower left corner and the Coulomb/insulator phase in the upper right corner.}
\label{fracphase}
\end{figure}

When $d\neq1$ the change of variables in Eq.~(\ref{shift}) can no longer be applied. Therefore the phase diagram in this case will be different from the $d=1$ case. We can determine the phase diagram by performing Monte Carlo simulations. As an example, the phase diagram for $d=3$ is shown in Fig.~\ref{fracphase}. Phase boundaries were determined from singularities in the specific heat. Note that in the Heisenberg model we can define a maximum of one hedgehog per lattice site, and so this model cannot easily be used to describe the binding of  multiple hedgehogs. Therefore all results for $d \neq 1$ come from the easy-plane $CP^1$ model for the spins. 

Figure~\ref{fracphase} presents the phase diagram in the variables $\lambda$ and $K$, with fixed $\beta=0.1$.
At small $\lambda$ and $K$, there is no energy cost for either hedgehogs or bosons, and they are independent. This leads to a paramagnet of spins and a superfluid of bosons. The $U(1)_{\rm spin}$ symmetry is preserved, and the $U(1)_{\rm boson}$ symmetry is broken. In contrast, at large $\lambda$ and $K$ both hedgehog and boson currents are forbidden, leading to a Coulomb phase of spins and an insulator of bosons. Here both the $U(1)_{\rm spin}$ and $U(1)_{\rm boson}$ symmetries are preserved, but the spin system has intrinsic topological order. 
At large $K$ but small $\lambda$, the Coulomb phase of the spin system survives and the hedghogs are gapped, but the bosons are condensed breaking the $U(1)_{\rm boson}$ symmetry.
On the other hand, at very small $K$ and large $\lambda$ the hedgehogs are proliferated and bosons are bound to them, and we are in the binding phase; here, neither symmetry is broken, and we will argue shortly that this is a Symmetry Enriched Topological (SET) phase. 

Note that similar phase diagram (at fixed small $\beta$) for $d=1$ contains only such four phases, where the binding phase is the SPT phase discussed in the previous sections. For $d=1$ these are the only phases, and due to the change of variables in Eq.~(\ref{shift}), the phase boundaries are all straight lines.
With $d\neq 1$, we have a new phase in the middle of the diagram, and no direct transition from the phase in the lower right to the binding phase. The middle phase can be understood as one in which hedgehog currents are proliferated, but their interactions are still too costly for objects with three hedgehogs and a boson to exist. Therefore such bound states are not proliferated, and individual bosons are also gapped.  We expect that this phase preserves the $U(1)$ symmetries from both the spins and bosons and is conventional paramagnet/insulator. The topological phase only arises when $K$ is lowered to the point that it does not penalize significantly objects with a hedgehog current of three and $\lambda$ is increased to strongly penalize any objects other than the $(Q, J) = (3, 1)$ bound states; at this point these bound states can form and proliferate, and the system enters the topological phase. For other values of $d$, the phase diagram is expected to have a similar form, with the exception of $d=2$, where in our studies of small system sizes the middle phase is not observed and there is a line of phase transitions between the lower left and upper right phases.

Let us now focus on the properties of the binding phase at large $\lambda$ and small $K$, which binds $d$ hedgehogs to a boson. This phase is distinct from the topological phase discussed earlier in this work. In particular, it has intrinsic topological order. Condensing bound states of $d$ hedgehogs causes the electric field lines in the phase to fractionalize, i.e., it is possible to have electric field lines of strength $1/d$. These fractionalized electric field lines are one of the gapped excitations of the system. The other elementary gapped excitation is a single hedgehog, which binds a boson charge of $1/d$. The hedgehog has well-defined statistics as it encircles the electric field lines, and we expect it to acquire a phase of $2\pi/d$ when this happens around the elementary fractionalized line. The matter fields $z_\uparrow$ and $z_\downarrow$ are confined, but still act as sources and sinks for the electric field lines of integer strength, therefore the line topological excitations in the system are only defined up to an integer, and we can say that the system has $\mathbb{Z}_d$ topological order.\cite{GukovKapustin,FracFaraday} 

In the Appendix we formally demonstrate the above properties by first removing the spinon matter fields and considering a CQED$\times U(1)_{\rm boson}$ system in which the monopoles of the compact electrodynamics are bound to bosons. Such CQED$\times U(1)_{\rm boson}$ models allow changes of variables similar to those possible in $U(1)\times U(1)$ models demonstrating SPT and SET phases of bosons in two dimensions\cite{FQHE}, which allow their properties to be readily determined. After the change of variables has been performed, we can couple the additional spinon matter fields to the CQED sector, and this gives us precisely the $CP^1\times U(1)_{\rm boson}$ model studied in this paper.

\begin{figure}
\includegraphics[width=0.75\linewidth,angle=-90]{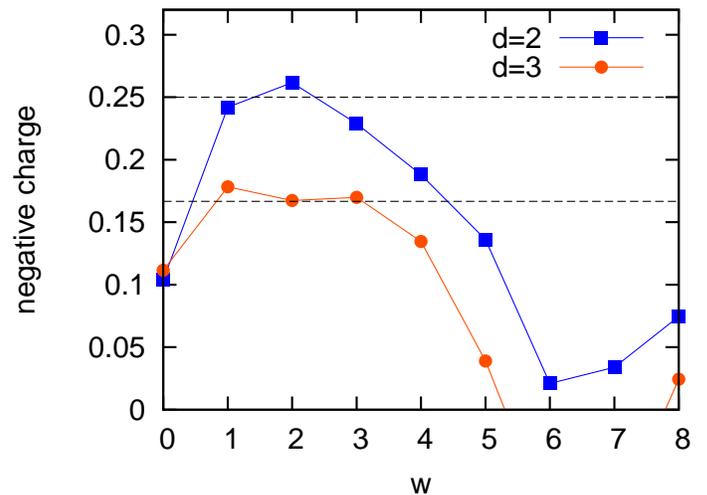}
\caption{Witten effect measurement, similar to Fig.~(\ref{witten}), but for the case where $d$ hedgehogs are bound to each boson. The amount of bound charge is given by $1/(2d)$, as expected (indicated by dashed lines). Parameters $K$, $\beta$ and $\lambda$ were chosen to put the bulk into the binding phase, while $\gamma$ was chosen to minimize the amount of charge away from $\scripty{r}=2$, though the value at $\scripty{r}=2$ is independent of this choice.
}
\label{wittend}
\end{figure}

The Witten effect and Hall effect measurements can be extended to the cases with multiple hedgehogs. For the Witten effect, the amount of bound charge will be modified, since now for each hedgehog there is a boson charge of $1/d$. Therefore the screening cloud will have a charge of $1/(2d)$. We have studied the cases of $d=2, 3$ in Monte Carlo and our results, shown in Fig.~\ref{wittend}, confirm this expectation. Recall that when measuring the Witten effect we used a ``biasing'' term to break degeneracy between positive and negative internal monopoles. This biasing term may also introduces some excess internal monopole or boson charge at the location of the external monopole. This excess is screened by the surrounding system. In the $d=1$ case, we found that for most values of $\gamma$, such as those shown in Fig.~\ref{diffgamma}, the screening length is quite short and so the Witten effect can still be clearly observed. In the case of $d>1$, the screening length seems to be much larger, which can lead to fluctuations of charge larger than the Witten effect we are trying to observe. We have chosen the biasing parameter in Fig.~\ref{wittend} in such a way as to minimize these fluctuations. At other values of $\gamma$ the Witten effect can still be observed but the observation is less clear due to these fluctuations.

\begin{figure}
\includegraphics[width=\linewidth]{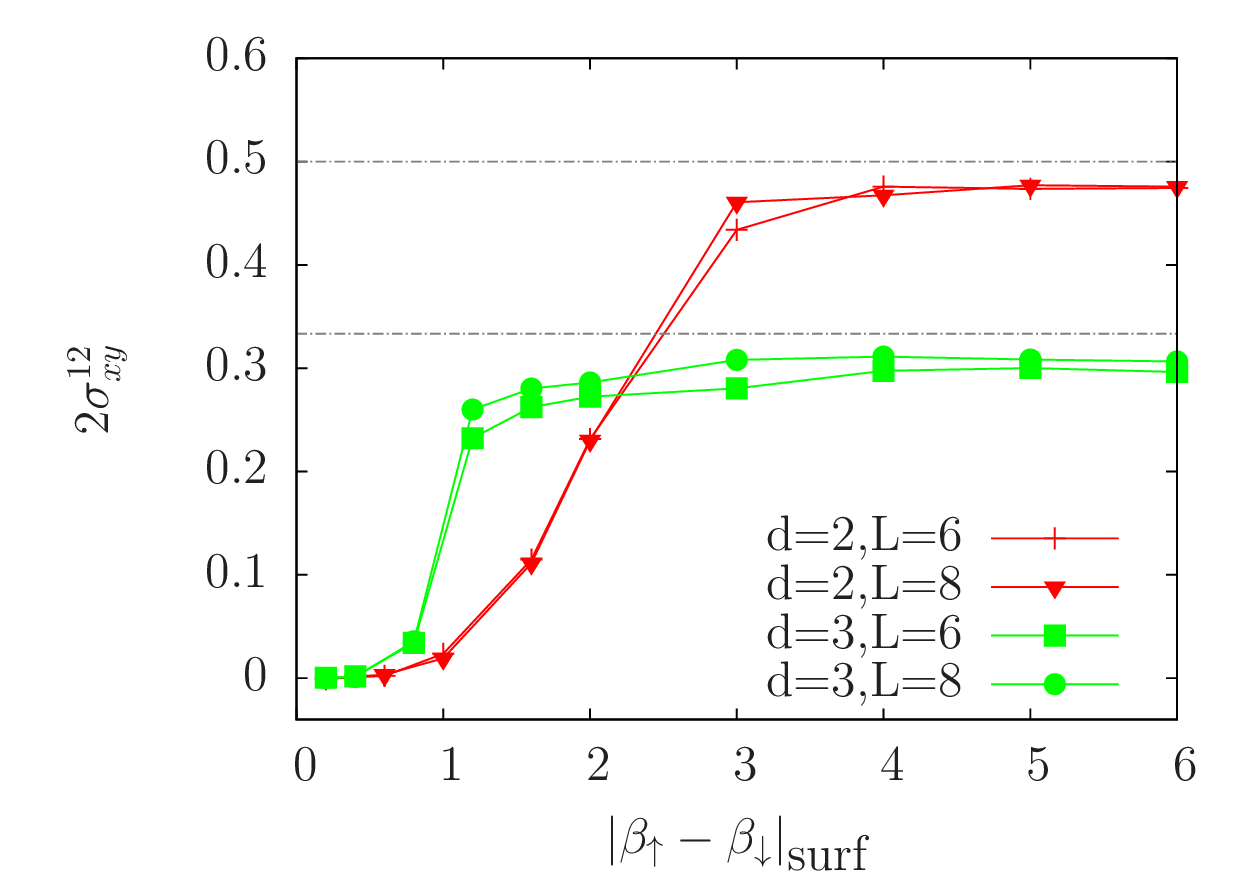}
\caption{Surface Hall conductivity at a boundary of an SET phase for diffent values of $d$, in units of $e^2/h$. We see that the Hall conductivity is given by $1/d$. Both surfaces have been averaged over to improve statistics. Dashed lines are drawn at $1/d$ to guide the eye. Data was taken with $K=0.2$, $\lambda=8$. Different values of $\beta_{\rm bulk}$ were used for different values of $d$, since the phase diagram changes as $d$ changes (see for example Fig.~\ref{fracphase}) and $\beta_{\rm bulk}$ needs to be chosen so that the system is in the topological phase.
}
\label{halldiff}
\end{figure}

We have also measured the surface Hall effect upon breaking the $\ztwot$ symmetry on the surface by applying the Zeeman field as in Sec.~\ref{cp1Hall}.
Our results for the Hall conductivity are shown in Fig.~\ref{halldiff}. We find that the surface Hall conductivity is given by $1/d$, which is one-half of the value found for a two-dimensional bosonic fractional quantum Hall effect.\cite{FQHE}  We can again rationalize this observation by considering a slab of the binding phase as in Fig.~\ref{monopoles} with the opposite Zeeman fields on the two surfaces, cf.\ discussion in the paragraph preceding Eq.~(\ref{Vmn}).

\section{Discussion and Conclusions}
\label{sec::discussion}
In this work we have realized a three-dimensional bosonic topological insulator in a lattice model which can be studied in Monte Carlo. The model works by binding bosons to point topological defects (hedgehogs). We determined the phase diagram in both the bulk and on the surface of the model. We were able to numerically extract signatures of the topological behavior: In the bulk of the model we observed a Witten effect, while on the surface with broken time-reversal symmetry we found a quantized Hall conductance with values distinct from those possible in a purely (2+1)D system. We also found other surface properties consistent with the bosonic TI, including a direct transition between surface phases with different broken symmetries, and a surface phase which breaks no symmetries and may possess topological order of a kind that would break $\ztwot$ in a purely (2+1)D system. Finally, we can also realize phases with intrinsic topological order in the bulk by binding multiple topological defects to each boson.

Our model can in principle be used to extract other properties of the bosonic topological insulator. One possible future direction is to determine the properties of the surface phase transitions, especially the transition between the two different surface superfluids. Another direction would be to find direct evidence of the exotic properties of the surface superfluids and surface topologically ordered phase. Such surface phases have generated much recent interest, as their excitations are expected to have properties not possible in a purely two-dimensional system.\cite{SenthilVishwanath,Chen2014,Cho2014} It would also be interesting to investigate the surface physics of the SET phases discussed in Sec.~\ref{section::multiple}.

In an earlier work on the two-dimensional bosonic topological phases,\cite{FQHE} we were also able to reconstruct (starting from Euclidean space-time actions) explicit microscopic Hamiltonians which realized the bosonic integer and fractional quantum Hall effects. We have been unable to do the same in the present three-dimensional case, but this would be a very interesting result. More broadly, the idea of binding bosons to topological defects may continue to yield precise models of bosonic topological phases.

The model in the main text can be thought of as having either $U(1)\rtimes\ztwot$ or $U(1)\times\ztwot$ symmetry and in both cases we are realizing the same phase.
Based on the cohomology classification\cite{WenScience,WenPRB}, the symmetry $U(1)\rtimes\ztwot$ has a $\ztwo^2$ classification in three dimensions; i.e.~there are two base phases with non-trivial topology, but one of the base phases comes from $\ztwot$ only.  There is only one phase that involves $U(1)$ in a non-trivial manner, and it is this phase that we are realizing in our construction and its signature is the Witten effect, which we discuss and observe in Sec.~\ref{subsec::witten}. Our construction cannot access the phase that comes from the $\ztwot$ symmetry only because we require the $U(1)$ symmetry.
In the $U(1) \times \ztwot$ case, the cohomology classification gives $\ztwo^3$, where one of the base phases is again from the $\ztwot$ alone, while the other two base phases involve $U(1)$ in a non-trivial manner.  Of the latter two, again we are realizing the one which has the statistical Witten effect as its signature.  The other base phase does not have the Witten effect, its signature is that the monopole is the Kramers doublet under time reversal.\cite{SenthilVishwanath, BiRasmussenXu}  Our model in principle has more symmetries and could be also deformed to produce this other phase. We do not consider this here but it is a possible direction for future work.  We are also not considering beyond cohomology phases, which bring yet another phase due to $\ztwot$ only in either case.\cite{SenthilVishwanath, Kapustin2014}.

The Appendix contains a more abstract compact quantum electrodynamics (CQED) model where multiple hedgehogs are bound to a boson, which realizes generalizations of SPT and SET phases in more abstract settings where lattice gauge theories are included as microscopic degrees of freedom. Such generalizations have been discussed formally recently,\cite{KapustinThorngren, McGreevy, KeyserlingkBurnell2014,GukovKapustin} and similar ideas can be useful for constructing explicit models and analysing physical properties of such {phases. As a simple demonstration, in a forthcoming publication we will consider a lattice CQED model where multiple monopoles condense and lead to a novel topological phase of CQED with fractionalized Faraday lines.\cite{FracFaraday}

\acknowledgments
We would like to thank M.~Metlitski, M.~P.~A.~Fisher, F.~Burnell, A.~Kapustin, C.~von~Keyserlingk, J.~Preskill, T.~Senthil, and A.~Vishwanath for many useful discussions.  We are particularly indebted to Max~Metlitski and Matthew~Fisher for sharing their unpublished results and crucial insights guiding our work, and also for carefully reading the manuscript and suggesting many improvements.
This research is supported by the National Science Foundation through grant DMR-1206096, and by the Caltech Institute of Quantum Information and Matter, an NSF Physics Frontiers Center with support of the Gordon and Betty Moore Foundation.

\appendix
\section{SPT- and SET-like phases in a CQED$\times$boson model in (3+1)D}
Here we consider a model whose degrees of freedom are compact quantum electrodynamics (CQED) residing on a (3+1)-dimensional cubic space-time lattice labelled by $R$, and bosons residing on a lattice labeled by $r$.  Topological excitations in the CQED are monopoles, which are quantum particles in three spatial dimensions, and our model realizes condensation of bound states of $d$ monopoles and $c$ bosons.  We will argue that when $d=1$, these phases are analogs of Bosonic Symmetry-Protected Topological phases for such CQED$\times$boson systems and have quantized cross-transverse response given by the integer $c$.  On the other hand, when $d > 1$ these phases are analogs of Bosonic Symmetry-Enriched Topological phases and have fractional cross-transverse response given by a rational number $c/d$; these phases also feature fractionalized Faraday line excitations of the CQED and fractionalized boson particle excitations, as well as non-trivial mutual statistics between the particle and line excitations.\cite{GukovKapustin,FracFaraday}  

While the CQED$\times$boson setting may appear artificial, this problem is relevant to the problem of SPT and SET phases of bosons in (3+1)D.  Specifically, in the main text we had a model with $U(1)\times U(1)$ symmetry, and we represented the first $U(1)$ system as an easy-plane \cp model, which has two matter fields (``spinons'') coupled to a compact gauge field.
We then considered binding of the monopoles in the compact gauge field and physical bosons of the second $U(1)$ symmetry.  The crucial difference with the CQED$\times$boson model is the additional matter fields present in the $CP^1\times$boson model.  We will compare the systems without and with such matter fields and will argue that the matter fields destroy the distinctions of the former model, except when protected by additional discrete symmetries.

Our CQED$\times$boson model is written in terms of compact gauge fields for the CQED part and integer-valued conserved currents for the boson part:
\begin{widetext}
\begin{eqnarray}
&& Z[\hext_{\mu\nu}(R), \Aext_\rho(r)] = {\sum_{\cJ_\rho(r)}}^\prime \int_0^{2\pi} Da_\mu(R) \int_0^{2\pi} \prod_{\mu < \nu} d\gamma_{\mu\nu} \sum_{\uu_{\mu\nu}(R) = -\infty}^\infty
e^{-S[a_\mu(R), \gamma_{\mu\nu}, \uu_{\mu\nu}(R), \cJ_\rho(r);~ \hext_{\mu\nu}(R), \Aext_\rho(r)]} ~,\label{A1}\\
&& S \!=\! \frac{K}{2} \sum_{R, \mu < \nu} \! \left[\omega_{\mu\nu}(R) - \hext_{\mu\nu}(R) - 2\pi \uu_{\mu\nu}(R) \right]^2 
+ \frac{\lambda}{2} \sum_{r, \rho} \left[c \cQ_\rho(r) + c \frac{\EuFrak{g}^{\rm ext}_\rho(r)}{2\pi} - d \cJ_\rho(r) \right]^2 
\! + i \sum_{r, \rho} \cJ_\rho(r) \Aext_\rho(r) , ~~ \label{Sorig} \\
&& \omega_{\mu\nu}(R) \equiv (\nabla_\mu a_\nu - \nabla_\nu a_\mu)(R) - \delta_{R_\mu = 0} \delta_{R_\nu = 0} \gamma_{\mu\nu} ~.
\end{eqnarray}
\end{widetext}
The above action can be compared to the action of Eqs.~(\ref{sspin}) and (\ref{cdbind}) in the main text. Compared to the main text, the above action is missing the $\beta$ term which couples the gauge field to ``spinons''. We also allow the option of binding multiple bosons---here binding $c$ bosons and $d$ monopoles. The boson sector has conserved particles and is coupled to a probing field $\Aext_\rho$ in the standard way, as in the main text. Unlike the main text, the CQED sector has conserved electric field lines and is coupled to a probing rank-2 field $\hext_{\mu\nu}(R)$.  In the absence of the binding term, the coupling of the CQED sector to $\hext_{\mu\nu}$ is standard;\cite{PolyakovBook} the additional piece in the binding term, while not important for much of the discussed long-distance physics, is the correct form keeping together the ``Dirac string'' $\uu_{\mu\nu}$ and the probing field $\hext_{\mu\nu}$. Note that in the main text the additional matter fields destroy the conservation of the electric field lines, and therefore the CQED sector cannot be probed by such an external rank-2 field.
Variables $\gamma_{\mu\nu}$ realize specific ``fluctuating boundary conditions'' in the compact gauge field variables; in a representation in terms of an integer-valued electromagnetic field tensor, this corresponds to requiring zero total field for each component.  Such details of the boundary conditions are not important for the bulk properties but are nice for a precise mathematical treatment in a system with periodic connectedness assumed here.
From variables $B_{\mu\nu}(R)$, we define the monopole four-current $\cQ_\rho(r)$ as in Eq.~(\ref{mondef}) in the main text.
We similarly define the four-vector $\EuFrak{g}^{\rm ext}_\rho(r)$ from $\hext_{\mu\nu}(R)$:
\begin{equation}
\EuFrak{g}^{\rm ext}_\rho(r) \equiv  \frac{1}{2} \epsilon_{\rho\sigma\mu\nu} \nabla_\sigma \hext_{\mu\nu} ~.
\end{equation}
Note that thus defined $\cQ_\rho(r)$ are conserved currents satisfying $\sum_\rho \nabla_\rho \cQ_\rho(r) = 0$, and they also satisfy the condition of zero total current in all directions: $\cQ_{{\rm tot}, \rho} \equiv \sum_r \cQ_\rho(r) = 0$.

For the boson sector, we use a representation in terms of integer-valued conserved currents $\cJ_\rho(r)$, which satisfy $\sum_\rho \nabla_\rho \cJ_\rho(r) = 0$.  We also require that the total boson current is equal to zero for all directions, $\cJ_{{\rm tot}, \rho} = 0$.  The primed sum over $\cJ_\rho(r)$ in the partition sum signifies all such constraints.  The condition of zero total current is again convenient for precise treatment on finite systems [namely, for performing change of variables involving $\cJ$ and $\cQ$ currents, Eq.~(\ref{SL2Z}) below], while for bulk properties one can ignore these details.  The ``binding'' term parametrized by $\lambda$ is the key interaction in the action Eq.~(\ref{Sorig}) and wants to have bound states of $d$ monopoles and $c$ bosons when $\lambda$ is large [i.e., it is minimized when $(\cQ, \cJ) = (d, c) \times {\rm integer}$].  

We first separate out the monopoles in the CQED sector as follows:
\begin{widetext}
\begin{eqnarray}
\!\!\!\!\! \sum_{\uu_{\mu\nu}(R) = -\infty}^\infty \!\!\!\!\!\!\!\!\! [\dots ] \!=\!\!\!\!\!
{\sum_{\cQ_\rho(r) = \frac{1}{2} \epsilon_{\rho\sigma\mu\nu} \nabla_\sigma \uu_{\mu\nu}^{(0)}}}^\prime \sum_{V_\mu(R) = -\infty}^\infty \sum_{\MM_{\mu\nu} = -\infty}^\infty 
\!\!\!\!\!\![\uu_{\mu\nu}(R) = \uu_{\mu\nu}^{(0)}(R) + (\nabla_\mu V_\nu - \nabla_\nu V_\mu)(R) + \delta_{R_\mu = 0} \delta_{R_\nu = 0} \MM_{\mu\nu} ]~.
\label{umn}
\end{eqnarray}
\end{widetext}
Here $\uu_{\mu\nu}^{(0)}(R)$ is an integer-valued field whose monopolicity gives $\cQ_\rho(r)$ and is treated as a fixed function of $\cQ_\rho(r)$.  $V_\mu(R)$ and $\MM_{\mu\nu}$ are independent integer-valued fields, where the latter appear from careful treatment of the boundary conditions.  Schematically, the above arises by dividing all configurations of $\uu_{\mu\nu}(R)$ into classes defined by $\cQ_\rho(r)$ and establishing how to recover all members of a given class from one representative.  Furthermore, any $\cQ_\rho(r)$ satisfying $\sum_\rho \nabla_\rho \cQ_\rho(r) = 0$ and $\cQ_{{\rm tot}, \rho} = 0$ can be represented as above using some integer-valued $\uu_{\mu\nu}(R)$, so the primed sum over $\cQ_\rho(r)$ can be viewed as signifying these constraints.  

A subtle point in the above is the redundancy in $V_\mu(R)$.  One way to address it precisely is as follows.  We can argue that from all links of the (3+1)D hyper-cubic lattice, we can select a subset of links such that we can take $V_\mu(R)$ as independent integer-valued variables on these links, while $V_\mu(R) = 0$ on all the other links.  We can also argue that the original CQED sector in Eq.~(\ref{A1}) can be equivalently formulated using compact gauge fields $a_\mu(R)$ that are non-zero only on exactly the same links as the independent $V_\mu(R)$.  We will assume this implicitly in all manipulations below, both for these variables and for the related variables $V_\mu^\prime(R)$, $\tilde{V}_\mu(R)$, $v_\mu(R)$, $k_{\mu}(R)$, and $\tilde{a}_\mu(R)$ that will appear below. 

We now operate with the constrained sums over the integer-valued currents $\cQ_\rho(r)$ and $\cJ_\rho(r)$, taken to be ``outside-most'' sums in the partition sum.  We change to new independent summation variables on each link:\cite{Gen2Loops}
\begin{equation}
\begin{array}{c}
\cP = a \cQ - b \cJ, \\
\cG = c \cQ - d \cJ,
\end{array} 
\leftrightarrow
\begin{array}{c}
\cQ = d \cP - b \cG, \\
\cJ = c \cP - a \cG ~.
\end{array}
\label{SL2Z}
\end{equation}
Here $c$ and $d$ are the same integers as in the binding term in Eq.~(\ref{Sorig}), while $a$ and $b$ are new integers such that $ad - bc = 1$, which makes the above transformation invertible in $\mathbb{Z}$.  If $c$ and $d$ are mutually prime, which we will assume throughout, we can always find such $a$ and $b$, while the arbitrariness $a \to a + kc, b \to b + kd$ does not affect any physical properties discussed below.

The new variables $\cP_\rho(r)$ and $\cG_\rho(r)$ are also conserved integer-valued currents with zero total currents.
For each $\cP_\rho(r)$ and $\cG_\rho(r)$, we can uniquely determine $\cQ_\rho(r)$ and hence the corresponding fixed $\uu_{\mu\nu}^{(0)}(R)$.  Let us also find for each $\cP_\rho(r)$ and $\cG_\rho(r)$ some fixed integer-valued $\uu_{P, \mu\nu}^{(0)}(R)$ and $\uu_{G, \mu\nu}^{(0)}(R)$ such that
\begin{equation}
\cP_\rho(r) = \frac{1}{2} \epsilon_{\rho\sigma\mu\nu} \nabla_\sigma \uu_{P, \mu\nu}^{(0)} ~, \quad 
\cG_\rho(r) = \frac{1}{2} \epsilon_{\rho\sigma\mu\nu} \nabla_\sigma \uu_{G, \mu\nu}^{(0)} ~.
\label{uPuG}
\end{equation}
We can guarantee that given the constraints satisfied by $\cP_\rho(r)$ and $\cG_\rho(r)$, such two-forms $\uu_{P, \mu\nu}^{(0)}(R)$ and $\uu_{G, \mu\nu}^{(0)}(R)$ always exist; while there are many possible choices, it does not matter which one we use as long as it stays fixed. 
We then have
\begin{equation}
\epsilon_{\rho\sigma\mu\nu} \nabla_\sigma \left[\uu_{\mu\nu}^{(0)} - d \uu_{P, \mu\nu}^{(0)} + b \uu_{G, \mu\nu}^{(0)} \right] = 0 ~,
\end{equation}
which implies that there exist integer-valued $V^\prime_\mu(R)$, $\MM^\prime_{\mu\nu}$ such that
\begin{eqnarray}
\uu_{\mu\nu}^{(0)}(R) &=& d \uu_{P, \mu\nu}^{(0)}(R) - b \uu_{G, \mu\nu}^{(0)}(R) \\
&+& (\nabla_\mu V^\prime_\nu - \nabla_\nu V^\prime_\mu)(R) + \delta_{R_\mu = 0} \delta_{R_\nu = 0} \MM^\prime_{\mu\nu} ~.\nonumber
\end{eqnarray}
We can again take $V^\prime_\mu(R)$, $\MM^\prime_{\mu\nu}$ as fixed functions of $\cP_\rho(r)$ and $\cG_\rho(r)$.  Now we can express $\uu_{\mu\nu}(R)$ in Eq.~(\ref{umn}) as
\begin{eqnarray}
\uu_{\mu\nu}(R) &=&  d \uu_{P, \mu\nu}^{(0)}(R) - b \uu_{G, \mu\nu}^{(0)}(R) \\
&+& (\nabla_\mu \tilde{V}_\nu - \nabla_\nu \tilde{V}_\mu)(R) + \delta_{R_\mu = 0} \delta_{R_\nu = 0} \tilde{\MM}_{\mu\nu} ~,\nonumber
\end{eqnarray}
where we have changed the summation variables from $V_\mu(R)$, $\MM_{\mu\nu}$ to $\tilde{V}_\mu(R) = V_\mu(R) + V^\prime_\mu(R)$, $\tilde{\MM}_{\mu\nu} = \MM_{\mu\nu} + \MM^\prime_{\mu\nu}$.

Let us write
\begin{eqnarray}
\tilde{V}_\mu(R) = d v_\mu(R) + k_\mu(R) ,~~
\tilde{\MM}_{\mu\nu} = d \mm_{\mu\nu} + \ell_{\mu\nu} ,
\end{eqnarray}
where $v_\mu(R)$ and $\mm_{\mu\nu}$ are arbitrary integers, while $k_\mu(R)$ and $\ell_{\mu\nu}$ are integers $0, 1, \dots, d-1$. 
Upon simple grouping of terms, we now have
\begin{eqnarray}
&& \omega_{\mu\nu}(R) - \hext_{\mu\nu}(R) - 2\pi \uu_{\mu\nu}(R) = \\
&& d \left[\tilde{\omega}_{\mu\nu}(R) - \frac{1}{d} \hext_{\mu\nu}(R) + \frac{2\pi b}{d} \uu_{G, \mu\nu}^{(0)}(R) - 2\pi \tilde{\uu}_{\mu\nu}(R) \right] ~,~\nonumber
\end{eqnarray}
where we defined
\begin{eqnarray}
\tilde{\omega}_{\mu\nu}(R) &\equiv& (\nabla_\mu \tilde{a}_\nu - \nabla_\nu \tilde{a}_\mu)(R) - \delta_{R_\mu = 0} \delta_{R_\nu = 0} \tilde{\gamma}_{\mu\nu} ~,\nonumber \\
\tilde{a}_\mu(R) &\equiv& \frac{a_\mu(R)}{d} - \frac{2\pi k_\mu(R)}{d} ~,\label{atilde} \\
\tilde{\gamma}_{\mu\nu} &\equiv& \frac{\gamma_{\mu\nu}}{d} + \frac{2\pi \ell_{\mu\nu}}{d} ~,\nonumber \\
\tilde{\uu}_{\mu\nu}(R) &\equiv& \uu_{P, \mu\nu}^{(0)}(R) + (\nabla_\mu v_\nu - \nabla_\nu v_\mu)(R) \nonumber \\
&& + \delta_{R_\mu = 0} \delta_{R_\nu = 0} \mm_{\mu\nu} ~.\nonumber
\end{eqnarray}
Note that the integration over $a_\mu(R)$ from $0$ to $2\pi$ and summation over $k_\mu(R)$ from $0$ to $d-1$ effectively corresponds to integration over $\tilde{a}_{\mu}(R)$ from $0$ to $2\pi$.  The same holds for $\tilde{\gamma}_{\mu\nu}$.
We can also express $\cP_\rho(r)$ from Eq.~(\ref{uPuG}) as $\cP_\rho(r) = \frac{1}{2} \epsilon_{\rho\sigma\mu\nu} \nabla_\sigma \tilde{\uu}_{\mu\nu}$ and see that $\uu_{P, \mu\nu}^{(0)}(R)$, $v_\mu(R)$, and $\mm_{\mu\nu}$ only appear in the above combination that gives $\tilde{\uu}_{\mu\nu}(R)$.  Furthermore, similarly to Eq.~(\ref{umn}), the summation over constrained $\cP_\rho(r)$ and unconstrained $v_\mu(R)$ and $\mm_{\mu\nu}$ is equivalent to a summation over unconstrained $\tilde{\uu}_{\mu\nu}$.  The partition sum becomes
\begin{widetext}
\begin{eqnarray}
&& Z[\hext_{\mu\nu}(R), \Aext_\rho(r)] = {\sum_{\cG_\rho(r)}}^\prime \int_0^{2\pi} D\tilde{a}_\mu(R) \int_0^{2\pi} \prod_{\mu < \nu} d\tilde{\gamma}_{\mu\nu} \sum_{\tilde{\uu}_{\mu\nu}(R) = -\infty}^\infty 
e^{-S[\tilde{a}_\mu(R), \tilde{\gamma}_{\mu\nu}, \tilde{\uu}_{\mu\nu}(R), \cG_\rho(r);~ \hext_{\mu\nu}(R), \Aext_\rho(r)]} ~,\\
&& S = \frac{K d^2}{2} \sum_{R, \mu < \nu} \left[\tilde{\omega}_{\mu\nu}(R) - \frac{1}{d} \hext_{\mu\nu}(R) + \frac{2\pi b}{d} \uu_{G, \mu\nu}^{(0)}(R) - 2\pi \tilde{\uu}_{\mu\nu}(R) \right]^2 
+ \frac{\lambda}{2} \sum_{r, \rho} \left[\cG_\rho(r) + c \frac{\EuFrak{g}^{\rm ext}_\rho(r)}{2\pi} \right]^2 \label{SKlambda} \\
&& ~~~ + i \sum_{r, \rho} \left[c \left(\frac{1}{2} \epsilon_{\rho\sigma\mu\nu} \nabla_\sigma \tilde{\uu}_{\mu\nu} \right)(r) - a \cG_\rho(r) \right] \Aext_\rho(r) ~.\nonumber
\end{eqnarray}
\end{widetext}
In this reformulation, we have a new CQED system represented by the compact gauge fields $\tilde{a}_\mu(R) \in [0, 2\pi)$ [and fluctuating boundary conditions realized with $\tilde\gamma_{\mu\nu}\in [0,2\pi)$], and a new boson system represented by the conserved current variables $\cG_\rho(r)$.  We will now argue that these new variables give a direct representation of gapped excitations in the topological phase obtained for small $K$ and large $\lambda$.  

For a quick illustration, let us set $\Aext_\rho(r) = 0$ everywhere.  In this case, we can perform summation over $\tilde{\uu}_{\mu\nu}(R)$ independently on each placket and obtain the Villain cosine for the combination
\begin{eqnarray}
\!\!\!\!
\Theta_{\mu\nu}(R) \equiv \tilde{\omega}_{\mu\nu}(R) - \frac{1}{d} \hext_{\mu\nu}(R) + \frac{2\pi b}{d} \uu_{G, \mu\nu}^{(0)}(R) ~.
\end{eqnarray}
Such cosine terms represent dynamics (motions) of quantum lines whose segments are created by $e^{i \tilde{a}_\mu(R)}$.  From the coupling to $\hext_{\mu\nu}(R)$, we conclude that the new lines carry electric field strength $1/d$ of the original electric field unit.  On the other hand, from the appearance of $\uu_{G, \mu\nu}^{(0)}(R)$ in $\Theta_{\mu\nu}(R)$ we conclude that the new lines ``see'' each $\cG$ particle as if it is carrying $b/d$ of monopole charge of the new CQED system.  We can state this equivalently as follows:  When a $\cG$ particle is taken around such a new line, there is a phase of $2\pi b/d$, i.e., there is non-trivial mutual statistics between the new lines and new particles.  For sufficiently small $K$ and large $\lambda$, the quantum lines created by $e^{i \tilde{a}_\mu(R)}$ and the particles represented by $\cG_\rho(r)$ are clearly gapped and are the only excitations in this phase.

By working a bit harder and keeping track of the $\Aext$ gauge field, we can also show that the $\cG$ particles carry charge of $1/d$ with respect to $\Aext$ (i.e., we also have boson charge fractionalization in this phase if $d > 1$), and that the system has a quantized cross-transverse response characterized by a rational number $c/d$.
First, we note that the current which couples to $\Aext_\rho(r)$ can be written as
\begin{eqnarray*}
c \cP_\rho(r) - a \cG_\rho(r) = c \left[\cP_\rho(r) - \frac{b}{d} \cG_\rho(r) \right] - \frac{1}{d} \cG_\rho(r) \\
= c \left(\frac{1}{2} \epsilon_{\rho\sigma\mu\nu} \nabla_\sigma \left[\tilde{\uu}_{\mu\nu} - \frac{b}{d} \uu_{G, \mu\nu}^{(0)} \right]\right)(r) - \frac{1}{d} \cG_\rho(r) ~,
\end{eqnarray*}
where we used $ad - bc = 1$ and manipulated so as to separate out the same combination of $\tilde{\uu}_{\mu\nu}$ and $\uu_{G, \mu\nu}^{(0)}$ as in the $K$-term in Eq.~(\ref{SKlambda}).
Second, on each placket $R, \mu < \nu$ we go from summation over integers $\tilde{\uu}_{\mu\nu}(R)$ to integration over real values as follows:
\begin{equation*}
\sum_{\tilde{\uu}_{\mu\nu}(R) = -\infty}^\infty \!\!\!\!\!\!\!\! [\dots ]
\!=\! \int_{-\infty}^\infty \!\!\!\!\! d\tilde{\uu}_{\mu\nu}(R) \!\!\!\!\! \sum_{\tilde{F}_{\mu\nu}(R) = -\infty}^\infty \!\!\!\!\! e^{-i 2\pi \tilde{F}_{\mu\nu}(R) \tilde{\uu}_{\mu\nu}(R)} [\dots ],
\end{equation*}
which can be viewed as an intermediate step in a Poisson resummation from $\tilde{B}_{\mu\nu}(R)$ to $\tilde{F}_{\mu\nu}(R)$. 
Finally, we perform Gaussian integration over the $\tilde{\uu}_{\mu\nu}(R)$, which can be simplified, e.g., by changing to new integration variable $x_{\mu\nu}(R) \equiv 2\pi \tilde{\uu}_{\mu\nu}(R) - \frac{2\pi b}{d} \uu_{G, \mu\nu}^{(0)}(R) + \frac{1}{d} \hext_{\mu\nu}(R) - \tilde{\omega}_{\mu\nu}(R)$.
The resulting action is:
\begin{widetext}
\begin{eqnarray}
S &=& \frac{1}{2 K d^2} \sum_{R, \mu < \nu} \left[\tilde{F}_{\mu\nu}(R) + c \frac{(\epsilon_{\mu\nu\sigma\rho} \nabla_\sigma \Aext_\rho)(R)}{2\pi} \right]^2
+ i \sum_{R, \mu < \nu} \tilde{F}_{\mu\nu}(R) \tilde{\omega}_{\mu\nu}(R)
+ \frac{\lambda}{2} \sum_{r, \rho} \left[\cG_\rho(r) + c \frac{\EuFrak{g}^{\rm ext}_\rho(r)}{2\pi} \right]^2 
\label{S_FG_wtildea} \\
&+& i \sum_{R, \mu < \nu} \frac{2\pi b}{d} \tilde{F}_{\mu\nu}(R) \uu_{G, \mu\nu}^{(0)}(R)
- i \sum_{R, \mu < \nu} \frac{1}{d} \tilde{F}_{\mu\nu}(R) \hext_{\mu\nu}(R)
- i \sum_{r, \rho} \frac{1}{d} \cG_\rho(r) \Aext_\rho(r)
- i \sum_{r, \rho} \frac{c}{2\pi d} \Aext_\rho(r) \EuFrak{g}^{\rm ext}_\rho(r) ~.\nonumber
\end{eqnarray}
\end{widetext}
We see that the formally introduced $\tilde{F}_{\mu\nu}(R)$ can be interpreted as integer-valued electromagnetic field tensor variables conjugate to the compact $\tilde{a}_\mu(R)$ variables.  Integrating out $\tilde{a}_\mu(R)$ gives ``Maxwell equations in the absence of charges'', $\sum_\nu \nabla_\nu \tilde{F}_{\mu\nu}(R) = 0$, while integrating out $\tilde{\gamma}_{\mu\nu}$ gives the condition $\sum_R \delta_{R_\mu = 0} \delta_{R_\nu = 0} \tilde{F}_{\mu\nu}(R) = 0$, which is equivalent to zero total field for each component.  For small $K$, the quantum lines whose worldsheets are represented by $\tilde{F}_{\mu\nu}(R)$ are gapped, and for large $\lambda$ the quantum particles whose worldlines are represented by $\cG_\rho(r)$ are gapped, so we have managed to express the partition sum completely in terms of gapped excitations in this phase.  From the coupling of $\tilde{F}_{\mu\nu}$ to $\hext_{\mu\nu}$ we explicitly see that the new lines carry fraction $1/d$ of the unit electric field strength, while from the coupling of $\cG_\rho$ to $\Aext_\rho$ we see that the new particles carry charge $1/d$ of the original bosons.  The coupling between $\tilde{F}_{\mu\nu}$ and $\uu_{G, \mu\nu}^{(0)}$ encodes $2\pi b/d$ statistical interaction between the new elementary line and particle excitations.  Finally, the last term $i \frac{c}{4\pi d} \sum \epsilon_{\rho\sigma\mu\nu} \Aext_\rho \nabla_\sigma \hext_{\mu\nu}$ is a property of the ``vacuum'' in these variables and represents a kind of cross-transverse response in this phase, which we see is quantized to the rational number $c/d$ in appropriate units.

When $d=1$ we have a quantized response, but the quantum numbers of the gapped excitations are not fractionalized, and therefore we have an SPT-like phase characterized by the integer $c$. When $d>1$ the quantum numbers of gapped excitations are fractionalized, and we have intrinsic topological order and an SET-like phase.

The cross-transverse response is due to unusual surface states when the above phase borders a trivial phase ($d=1$, $c=0$). Analysis of the surface theory in Sec.~\ref{subsec::cp1surface} can be carried over to this case in the absence of the spinon matter, and we see that the surface of the SPT-like phase $(d=1,c=1)$ in the CQED$\times$boson system has an emergent non-compact electrodynamics whose flux couples to $\Aext$ as in Eq.~(\ref{surfaceCS}). This tells us that the gapless surface mode can be viewed as a (2+1)D photon. Equivalently, the surface can be viewed as a boson system without vortices, which is always in the spin-wave phase, and therefore the (2+1)D gapless mode can also be viewed as a (2+1)D phonon.

We can now make the connection to the $CP^1\times$boson model in the main text, which we will refer to as $U(1)_{\rm spin} \times U(1)_{\rm boson}$ model.  Here we have additional ``spinon'' matter fields coupled to the compact gauge field, schematically
\begin{eqnarray}
\delta S &=& S_{\rm spinons}[J_\up, J_\dn] + i \sum_{R, \mu} [J_{\up, \mu}(R) + J_{\dn, \mu}(R)] a_\mu(R),\nonumber
\end{eqnarray}
where $J_\up$ and $J_\dn$ are integer-valued spinon currents conjugate to the $\phi_\up$ and $\phi_\dn$ variables in Eq.~(\ref{sspin}). Using Eq.~(\ref{atilde}) and dropping all contributions of the form $i 2\pi \times$integer, we can write
\begin{eqnarray}
\delta S &=& S_{\rm spinons}[J_\up, J_\dn] + i d \sum_{R, \mu} [J_{\up, \mu}(R) + J_{\dn, \mu}(R)] \tilde{a}_\mu(R) ~.\nonumber
\end{eqnarray}
All manipulations leading to Eq.~(\ref{S_FG_wtildea}) do not touch the gauge field $\tilde{a}_\mu(R)$ and remain valid also in the presence of the spinon matter.  Now, since the original electric field lines have sources and sinks, the probing field $\hext_{\mu\nu}$ ceases to be meaningful and will be dropped; in particular, the characterization using rational $c/d$ for the cross-transverse response collapses.  On the other hand, the coupling of the new $\cG$ particles to $\Aext$ remains unaffected, and we see that they carry $1/d$ fractional charge with respect to $U(1)_{\rm boson}$.  Furthermore, upon integrating out the $\tilde{a}_\mu(R)$ fields, we now have
\begin{equation}
\sum_\nu \nabla_\nu \tilde{F}_{\mu\nu}(R) = -d [J_{\up, \mu}(R) + J_{\dn, \mu}(R)] ~,
\end{equation}
i.e., the spinons act as sources and sinks of the new field lines encoded by the integer-valued $\tilde{F}_{\mu\nu}(R)$, but these sources and sinks are in multiples of $d$.  We assume that the spinons are gapped, and they will clearly be confined in the regime of small $K$; however, they destroy the structure that we had of the integer-labeled closed lines.  Nevertheless, this destruction happens only for multiples of $d$, while it is still meaningful to talk about closed line excitations of strengths modulo $d$, i.e., we have line excitations labeled by $\mathbb{Z}_d$.  The final result is that there is no fractionalization of the $U(1)_{\rm spin}$, but we still have $\mathbb{Z}_d$ quantum line excitations and fractionalized particle excitations carrying charge $1/d$ with respect to the $U(1)_{\rm boson}$, with mutual statistics between these lines and particles. 

Note that in either the $d=1$ or $d>1$ case, when the system contains the $U(1)_{\rm spin}$ and $\ztwot$ symmetries and $c$ is odd, the resulting phase is distinct from a phase which is a direct product of a trivial spin state times a trivial (or $1/d$ fractionalized) boson state. This distinction can be observed by measuring a quantized Witten effect as discussed in the main text. To see this one also needs to carefully include the external gauge field coupled to $U(1)_{\rm spin}$ as we did in Eq.~(\ref{withA}), particularly in the presence of external monopoles, which we did not keep track of in this Appendix.

\bibliography{SO34D}
\end{document}